# Eigenvalue Distributions of Sums and Products of Large Random Matrices via Incremental Matrix Expansions

Matthew J. M. Peacock[†]

School of Electrical & Information Engineering

The University of Sydney, Sydney, NSW 2006, Australia.

mpeac@ee.usyd.edu.au

Iain B. Collings

CSIRO ICT Centre

Sydney, Australia.

iain.collings@csiro.au

Michael L. Honig[‡]

Department of Electrical & Computer Engineering

Northwestern University, Evanston, IL 60208, USA.

mh@ece.northwestern.edu

**Abstract**

This paper uses an incremental matrix expansion approach to derive asymptotic eigenvalue distributions (a.e.d.'s) of sums and products of large random matrices. We show that the result can be derived directly as a consequence of two common assumptions, and matches the results obtained from using R- and S-transforms in free probability theory. We also give a direct derivation of the a.e.d. of the sum of certain random matrices which are not free. This is used to determine the asymptotic signal-to-interference-ratio of a multiuser CDMA system with a minimum mean-square error linear receiver.

**Index Terms**

Large System, Free Probability, R-transform, S-transform, MMSE, CDMA.

[†] Matthew Peacock is supported in part by the Australian CSIRO

[‡] Supported by the U.S. Army Research Office under DAAD19-99-1-0288 and the National Science Foundation under grant CCR-0310809

## I. INTRODUCTION

Asymptotic analysis of linear multi-input multi-output communications systems has yielded significant insights into their performance and design (e.g., see [1] and references therein). In particular, asymptotic, or large-system, analysis of the minimum-mean-squared-error (MMSE) receiver has been studied extensively for code-division multiple-access (CDMA) systems, using results such as the Silverstein-Bai theorem [2], Girko's law [3], and free probability [4, 5].

Free probability is concerned with non-commutative random variables, of which asymptotically large random matrices are a canonical example. In free probability, the notion of independence (from commutative probability theory) is replaced by the notion of freeness. Methods for finding the asymptotic eigenvalue distribution (a.e.d.) of sums and products of free non-commutative random variables were developed by Voiculescu [6, 7], and apply what are known as the R- and S-transforms, respectively. Recently, free probability has been used to analyze several aspects of communications systems [8–13].

In this paper, we show that for asymptotically large random matrices, these sum and product a.e.d.s can be derived in a more direct manner than in [6, 7]. The derivations arise directly as a consequence of two conditions (which ensure freeness), and do not rely on non-commutative free probability theory. Instead, we apply an incremental matrix expansion approach [14], which is a generalization of the techniques used in [2, 15]. Similar, yet different derivations of these results can also be found in the mathematical physics literature [16, 17], however we believe the derivations found there are less accessible. These derivations help to explain key results from free probability theory, in particular, the R- and S-transforms.

The incremental matrix expansion approach can also be used to determine the a.e.d. of sums and products of certain *non-free* random matrices. This was demonstrated in [14], where we previously considered the large-system transient performance of adaptive least-squares receivers. In this paper, we extend the approach in order to consider certain multi-user CDMA systems in frequency-selective channels, which includes the single-cell multiple-signature-per-user uplink and the multiple-cell downlink. Specifically, we consider direct-sequence (DS) and multi-carrier (MC) CDMA systems, which are well known to be equivalent in the large system limit (see e.g., [18]), and as such we refer to the common model as DS/MC-CDMA. We previously presented an approximate solution to a special case of this problem in [18, 19], where a sum of non-free matrices is approximated by a sum of equivalent unitarily invariant matrices, which are asymptotically free. In this paper, we determine an exact solution to the problem, which is also significantly easier to compute than the approximate result in [18]. A special case of the solution is seen to agree with results in the mathematical literature [20]. Numerical examples show that the exact large-system results closely match simulated finite-system results.

## II. A.E.D.'S OF SUMS AND PRODUCTS OF UNITARILY INVARIANT MATRICES

In [14], we outlined an extension to the approach of [2] for computing the a.e.d. of certain types of large random matrices using elementary matrix operations. This approach gives the same results as would be obtained if results from free probability[1] were used, but the derivations are more direct. Here we show that the a.e.d.'s of sums and products of free random matrices can also be derived using this approach.

The a.e.d. of sums and products of asymptotically free random matrices can be computed, respectively, using the so-called R- and S-transforms from free probability, given the a.e.d. of each component term [6, 7]. This is analogous to the way the Fourier transform is used to compute the distribution of a sum of scalar independent random variables. As such, R- and S-transforms are often described as performing additive or multiplicative *free convolution* of the component distributions. In what follows, we will show that the sum and product distributions can be derived in a more direct manner, which does not explicitly require free probability results, but depends on two assumptions satisfied by canonical examples of free random matrices [5].

According to [5, Theorem 4.3.5], an independent family of $N \times N$ Hermitian random matrices[2] $(\mathbf{X}_j)_{j=1,\ldots,J}$ are almost surely asymptotically free as $N \to \infty$ provided that for each $j = 1, \ldots, J$:

**Assumption 1** $\mathbf{X}_j$ *is unitarily invariant. That is, the joint distribution of the matrix elements is invariant to left or right multiplication by unitary matrices.*

**Assumption 2** *The empirical distribution function (e.d.f.) of the eigenvalues of $\mathbf{X}_j$ almost surely converges in distribution to a compactly supported probability measure on $\mathbb{R}^*$ as $N \to \infty$.*

We shall define $X_j$ as a scalar random variable according to the a.e.d. of $\mathbf{X}_j$ for each $j = 1, \ldots, J$.

---

[1] A straightforward introduction to free probability can be found in [12, Section V], which also contains references to further information.

[2] **Notation:** All vectors are defined as column vectors and designated with bold lower case; all matrices are given in bold upper case; $(\cdot)^\dagger$ denotes Hermitian (i.e. complex conjugate) transpose; $(\cdot)^\ddagger$ denotes the operation $\mathbf{X}^\ddagger = \mathbf{X}\mathbf{X}^\dagger$; $\mathrm{tr}[\cdot]$ denotes the matrix trace; $|\cdot|$ and $\|\cdot\|$ denote the Euclidian and induced spectral norms, respectively; $\mathbf{I}_N$ denotes the $N \times N$ identity matrix; and, expectation is denoted $\mathbf{E}[\cdot]$.

## A. Ramifications of Assumptions 1–2

*1) Assumption 1 (Unitary Invariance):* Denote the singular value decomposition of $\mathbf{X}_j$ as $\mathbf{V}_j \mathbf{D}_j^2 \mathbf{V}_j^\dagger$. Due to the assumption of unitary invariance of $\mathbf{X}_j$, without loss of generality we may assume that $\mathbf{V}_j$ is independent $N \times N$ Haar[3] unitary. In what follows, $\mathbf{v}_{j,k}$ denotes the $k^{\text{th}}$ column of $\mathbf{V}_j$ and $D_{j,k}$ denotes the $k^{\text{th}}$ diagonal element of $\mathbf{D}_j^2$ for $1 \leq j \leq J$ and $1 \leq k \leq N$.

*2) Assumption 2 (Convergence of empirical distributions):* From [21], Assumption 2 implies

$$\lim_{N \to \infty} \frac{1}{N} \sum_{k=1}^{N} \mathrm{f}(D_{j,k}) = \mathbf{E}\left[\mathrm{f}(X_j)\right] \tag{1}$$

almost surely, where $f : \mathbb{R}^* \to \mathbb{R}^*$ is any (fixed) bounded continuous function on the support of $X_j$. Denote $D_{\max} = \max_{j \leq J} \sup_N \|\mathbf{X}_j\|$, which is finite due to Assumption 2.

We shall assume that the distribution of $X_j$, $j = 1 \ldots J$, is non-trivial, i.e., does not have all mass at zero, since we shall require $\mathbf{E}[X_j] > 0$.

## B. Sums of Unitarily Invariant Matrices

We wish to determine the a.e.d. (i.e., as $N \to \infty$) of $\sum_{j=1}^{J} \mathbf{X}_j$ where the $\mathbf{X}_j$ are Hermitian, $N \times N$ independent random matrices satisfying Assumptions 1 and 2. Equivalently, it is more convenient to determine the Stieltjés transform[4] of the distribution, which may then be inverted. That is, we seek $G_C(z) = \lim_{N \to \infty} G_C^N(z)$ where $G_C^N(z) = \frac{1}{N}\mathrm{tr}[\mathbf{C}^{-1}]$ and $\mathbf{C} = -z\mathbf{I}_N + \sum_{j=1}^{J} \mathbf{X}_j$. To simplify the proof, we shall also assume $|z| < \infty$.

Before we begin the formal derivation, we first explain the general method we use. Following the approach in [14], it can be seen that at some point, the matrix inversion lemma will be used to remove column $k$ of $\mathbf{V}_j$ from $\mathbf{C}$, which will give a term of the form

$$\mathbf{v}_{j,k}^\dagger \mathbf{C}_{j,k}^{-1} \mathbf{v}_{j,k} \tag{2}$$

where $\mathbf{C}_{j,k}$ is $\mathbf{C}$ with $\mathbf{v}_{j,k}$ removed from $\mathbf{X}_j$. Now, in order to analyze such a term asymptotically, typically the next step is to convert this quadratic form into an expression involving a matrix trace. Since $\mathbf{v}_{j,k}$ is a column from a Haar distributed matrix, the term in (2) can be evaluated asymptotically from Lemma 5 in Appendix I, which gives

$$\mathbf{v}_{j,k}^\dagger \mathbf{C}_{j,k}^{-1} \mathbf{v}_{j,k} \asymp \frac{1}{N - (N-1)} \mathrm{tr}[(\mathbf{I}_N - \mathbf{V}_j \mathbf{V}_j^\dagger + \mathbf{v}_{j,k}\mathbf{v}_{j,k}^\dagger)\mathbf{C}_{j,k}^{-1}] = \mathbf{v}_{j,k}^\dagger \mathbf{C}_{j,k}^{-1} \mathbf{v}_{j,k} \tag{3}$$

---

[3]A square random matrix $\mathbf{\Omega}$ is Haar distributed if its probability distribution is invariant to left or right multiplication by any constant unitary matrix.

[4]The Stieltjés (or Cauchy) transform of the distribution of a random variable $X \in \mathbb{R}^*$ is $\mathbf{E}[\frac{1}{X-z}]$, where $z \in \mathbb{C}^+$ is the transform variable, and $\mathbb{C}^+ = \{x \mid x \in \mathbb{C}, \mathrm{Im}(x) > 0\}$.

where $\asymp$ denotes "asymptotic equivalence", as defined in Definition 1 in Appendix I. Unfortunately, this tells us nothing new about (2). Therefore, we introduce an intermediate step in the derivations, where we reduce the rank of $\mathbf{V}_j$ to some value $K < N$, and consider the asymptotic limit $(N, K) \to \infty$ with $K/N \to \alpha$ where $\alpha \in (0, 1)$. The result we seek is then obtained by letting $\alpha$ go to unity from below, denoted $\alpha \to 1^-$.

Formally, we seek

$$G_C(z) = \lim_{\alpha \to 1^-} \acute{G}_C(z, \alpha) \tag{4}$$

where

$$\acute{G}_C(z, \alpha) = \lim_{\substack{(N,K) \to \infty \\ K/N \to \alpha}} \acute{G}_C^N(z, \alpha) \quad , \alpha \in (0, 1), \tag{5}$$

$$\acute{G}_C^N(z, \alpha) = \frac{1}{N} \mathrm{tr}\left[\acute{\mathbf{C}}^{-1}\right], \tag{6}$$

$$\acute{\mathbf{C}} = -z\mathbf{I}_N + \sum_{j=1}^{J} \acute{\mathbf{X}}_j, \tag{7}$$

$$\acute{\mathbf{X}}_j = (\acute{\mathbf{V}}_j \acute{\mathbf{D}}_j)^{\ddagger} \tag{8}$$

and $\acute{\mathbf{V}}_j$ contains the first $K < N$ columns of $\mathbf{V}_j$, and $\acute{\mathbf{D}}_j$ is the corresponding $K \times K$ submatrix of $\mathbf{D}_j$.

Following the incremental matrix expansion approach described in [14], the next step is to remove column $k$ of $\acute{\mathbf{V}}_j$ from $\acute{\mathbf{C}}$, i.e., $\acute{\mathbf{C}} = \acute{\mathbf{C}}_{j,k} + D_{j,k}\mathbf{v}_{j,k}\mathbf{v}_{j,k}^{\dagger}$. We have

$$\acute{\mathbf{C}}^{-1}\mathbf{v}_{j,k} = \frac{\acute{\mathbf{C}}_{j,k}^{-1}\mathbf{v}_{j,k}}{1 + D_{j,k}\rho_{j,k}^N} \tag{9}$$

from the matrix inversion lemma, where $\rho_{j,k}^N = \mathbf{v}_{j,k}^{\dagger} \acute{\mathbf{C}}_{j,k}^{-1} \mathbf{v}_{j,k}$. It is shown in Appendix II that

$$\max_{k \leq K} \left| \rho_{j,k}^N - \rho_j^N \right| \xrightarrow{a.s.} 0 \tag{10}$$

as $(N, K) \to \infty$ with $K/N \to \alpha \in (0, 1)$, where

$$\rho_j^N = \frac{1}{N - K} \mathrm{tr}[\boldsymbol{\Upsilon}_j \acute{\mathbf{C}}^{-1}], \tag{11}$$

$$\boldsymbol{\Upsilon}_j = \mathbf{I}_N - \acute{\mathbf{V}}_j \acute{\mathbf{V}}_j^{\dagger} \tag{12}$$

Expanding the identity $\mathbf{I}_N = \acute{\mathbf{C}}\acute{\mathbf{C}}^{-1}$ using (9) gives

$$1 = \frac{1}{N}\text{tr}[\acute{\mathbf{C}}\acute{\mathbf{C}}^{-1}] \tag{13}$$

$$= -z\acute{G}_C^N(z,\alpha) + \frac{1}{N}\sum_{j=1}^{J}\sum_{k=1}^{K} D_{j,k}\mathbf{v}_{j,k}^{\dagger}\acute{\mathbf{C}}^{-1}\mathbf{v}_{j,k} \tag{14}$$

$$= -z\acute{G}_C^N(z,\alpha) + \alpha J - \alpha\sum_{j=1}^{J}\frac{1}{K}\sum_{k=1}^{K}\frac{1}{1+D_{j,k}\rho_{j,k}^N} \tag{15}$$

and similarly, we have from (11)

$$\rho_j^N = \frac{1}{1-\alpha}\left(\acute{G}_C^N(z,\alpha) - \alpha\frac{1}{K}\sum_{k=1}^{K}\frac{\rho_{j,k}^N}{1+D_{j,k}\rho_{j,k}^N}\right) \tag{16}$$

We now concentrate on a realization for which (1) and (10) holds. Under this assumption, it is shown in Appendix II that

$$\max_{k\leq K}\left|\frac{1}{1+D_{j,k}\rho_{j,k}^N} - \frac{1}{1+D_{j,k}\rho_j^N}\right| \to 0 \tag{17}$$

$$\max_{k\leq K}\left|\frac{\rho_{j,k}^N}{1+D_{j,k}\rho_{j,k}^N} - \frac{\rho_j^N}{1+D_{j,k}\rho_j^N}\right| \to 0 \tag{18}$$

Moreover, due to Lemma 7 in Appendix I and (17)–(18), we have

$$\left|\frac{1}{K}\sum_{k=1}^{K}\left(\frac{1}{1+D_{j,k}\rho_{j,k}^N} - \frac{1}{1+D_{j,k}\rho_j^N}\right)\right| \to 0 \tag{19}$$

$$\left|\frac{1}{K}\sum_{k=1}^{K}\left(\frac{\rho_{j,k}^N}{1+D_{j,k}\rho_{j,k}^N} - \frac{\rho_j^N}{1+D_{j,k}\rho_j^N}\right)\right| \to 0 \tag{20}$$

It now follows from (1), (10), (15), (16), (19), and (20) that $\left|\acute{G}_C^N(z,\alpha) - \acute{G}_C(z,\alpha)\right| \to 0$ and $\left|\rho_j^N - \rho_j\right| \to 0$, $j=1,\ldots,J$, where $\acute{G}_C(z,\alpha)$ and $\rho_j$, $j=1,\ldots,J$, satisfy:

$$\acute{G}_C(z,\alpha) = -z^{-1}\left(1 - \alpha J + \alpha\sum_{j=1}^{J}\mathcal{E}_j^{\text{A}}\right) \tag{21}$$

$$\rho_j = \frac{\acute{G}_C(z,\alpha)}{\alpha(\mathcal{E}_j^{\text{A}} - 1) + 1} \tag{22}$$

and $\mathcal{E}_j^{\text{A}} = \mathbf{E}\left[\frac{1}{1+X_j\rho_j}\right]$.

As discussed at the beginning of this section, taking $\alpha \to 1^-$ in (21) and (22), we obtain the $J+1$ simultaneous equations in the variables $G_C(z)$, $\rho_j$, $j=1,\ldots,J$, given by

$$G_C(z) = \frac{J-1}{z + \sum_{j=1}^{J}\rho_j^{-1}} \tag{23}$$

$$G_C(z) = \mathbf{E}\left[\frac{1}{X_j + \rho_j^{-1}}\right] \quad, \quad j=1,\ldots,J \tag{24}$$

Since a solution to these equations exists and is unique [22, Theorem 2.1], we have that $G_C^N(z) \to G_C(z)$ with probability 1.

Finally, it can easily be verified that this result matches that obtained using the R-transform from free probability theory [6]. We emphasize that this derivation does not explicitly rely on free probability results; rather it relies only on Assumptions 1–2.

## C. Products of Unitarily Invariant Matrices

We wish to determine the a.e.d. (i.e., as $N \to \infty$), of $\prod_{j=1}^{J} \mathbf{X}_j$ where the $\mathbf{X}_j$ are as defined previously. Equivalently, it is more convenient to determine the Stieltjés transform of the distribution. That is, we seek $G_B^N(z) = \lim_{N \to \infty} \frac{1}{N} \text{tr}[\mathbf{B}^{-1}]$, where $\mathbf{B} = -z\mathbf{I}_N + \prod_{j=1}^{J} \mathbf{X}_j$. In what follows, we assume $J = 2$ in order to simplify the derivations. Of course, the result may be applied recursively to obtain the a.e.d. for $J > 2$. Also, to simplify the proof, we shall assume that $|z| < \infty$.

As explained in Section II-B, rather than attempting to derive $G_B^N(z)$ directly, we consider an associated problem where the rank of $\mathbf{V}_j$, $j = 1, \ldots, J$, is reduced to $K$. We then take the asymptotic limit $(N, K) \to \infty$ with $K/N \to \infty$, and obtain the desired solution by taking $\alpha$ to unity from below. That is, we define

$$G_B(z) = \lim_{\alpha \to 1^-} \acute{G}_B(z, \alpha) \tag{25}$$

where

$$\acute{G}_B(z, \alpha) = \lim_{\substack{(N,K) \to \infty \\ K/N \to \alpha}} \acute{G}_B^N(z, \alpha) \quad , \alpha \in (0, 1), \tag{26}$$

$$\acute{G}_B^N(z, \alpha) = \frac{1}{N} \text{tr} \left[ \acute{\mathbf{B}}^{-1} \right], \tag{27}$$

$$\acute{\mathbf{B}} = -z\mathbf{I}_N + \acute{\mathbf{X}}_1 \acute{\mathbf{X}}_2 \tag{28}$$

and $\acute{\mathbf{X}}_i$ is defined in (8).

As in Section II-B, the next step is to remove column $k \leq K$ from $\acute{\mathbf{V}}_i$ within $\acute{\mathbf{B}}$. Define $\acute{\mathbf{B}}_{1,k} = \acute{\mathbf{B}} - D_{1,k}\mathbf{v}_{1,k}\mathbf{v}_{1,k}^\dagger \acute{\mathbf{X}}_2$ and $\acute{\mathbf{B}}_{2,k} = \acute{\mathbf{B}} - D_{2,k}\acute{\mathbf{X}}_1\mathbf{v}_{2,k}\mathbf{v}_{2,k}^\dagger$. From the matrix inversion lemma, we have

$$\mathbf{v}_{1,k}^\dagger \acute{\mathbf{X}}_2 \acute{\mathbf{B}}^{-1} \mathbf{v}_{1,k} = \frac{\pi_{1,k}^N}{1 + D_{1,k}\pi_{1,k}^N} \tag{29}$$

$$\mathbf{v}_{2,k}^\dagger \acute{\mathbf{B}}^{-1} \mathbf{X}_1 \mathbf{v}_{2,k} = \frac{\pi_{2,k}^N}{1 + D_{2,k}\pi_{2,k}^N} \tag{30}$$

$$\mathbf{v}_{2,k}^\dagger \acute{\mathbf{B}}^{-1} \mathbf{v}_{2,k} = \frac{-z^{-1}}{1 + D_{2,k}\pi_{2,k}^N} \tag{31}$$

where

$$\pi_{1,k}^N = \mathbf{v}_{1,k}^\dagger \acute{\mathbf{X}}_2 \acute{\mathbf{B}}_{1,k}^{-1} \mathbf{v}_{1,k} \tag{32}$$

$$\pi_{2,k}^N = \mathbf{v}_{2,k}^\dagger \acute{\mathbf{B}}_{2,k}^{-1} \acute{\mathbf{X}}_1 \mathbf{v}_{2,k} \tag{33}$$

and we have used $\mathbf{v}_{2,k}^\dagger \acute{\mathbf{B}}_{2,k}^{-1} \mathbf{v}_{2,k} = -z^{-1}$, which follows from $\acute{\mathbf{B}}_{2,k} \mathbf{v}_{2,k} = -z \mathbf{v}_{2,k}$.

It is shown in Appendix III that

$$\max_{k \leq K} \left| \pi_{j,k}^N - \pi_j^N \right| \xrightarrow{a.s.} 0 \tag{34}$$

for $i = 1, 2$, as $(N, K) \to \infty$ with $K/N \to \alpha \in (0, 1)$, where

$$\pi_1^N = \frac{1}{N-K} \text{tr} \left[ \mathbf{\Upsilon}_1 \acute{\mathbf{X}}_2 \acute{\mathbf{B}}^{-1} \right] \tag{35}$$

$$\pi_2^N = \frac{1}{N-K} \text{tr} \left[ \mathbf{\Upsilon}_2 \acute{\mathbf{B}}^{-1} \acute{\mathbf{X}}_1 \right] \tag{36}$$

and $\mathbf{\Upsilon}_j$ is defined in (12).

Expanding the normalized trace of the identity $\mathbf{I}_N = \acute{\mathbf{B}} \acute{\mathbf{B}}^{-1}$ using (29) and (30) we obtain

$$1 + z\acute{G}_B^N(z,\alpha) = \begin{cases} \frac{1}{N} \sum_{k=1}^K D_{1,k} \mathbf{v}_{1,k}^\dagger \acute{\mathbf{X}}_2 \acute{\mathbf{B}}^{-1} \mathbf{v}_{1,k} \\ \frac{1}{N} \sum_{k=1}^K D_{2,k} \mathbf{v}_{2,k}^\dagger \acute{\mathbf{B}}^{-1} \acute{\mathbf{X}}_1 \mathbf{v}_{2,k} \end{cases} \tag{37}$$

$$= \alpha - \frac{1}{N} \sum_{k=1}^K \frac{1}{1 + D_{j,k} \pi_{j,k}^N} \quad , j = 1, 2. \tag{38}$$

and similarly expanding (35) using (29)–(31) we obtain

$$\pi_1^N = \frac{\alpha}{1-\alpha} \left( -z^{-1} \frac{1}{K} \sum_{k=1}^K \frac{D_{2,k}}{1 + D_{2,k} \pi_{2,k}} - \frac{1}{K} \sum_{k=1}^K \frac{\pi_{1,k}}{1 + D_{1,k} \pi_{1,k}} \right) \tag{39}$$

We now concentrate on a realization for which (1) and (34) holds. Under this assumption, it is shown in Appendix III that

$$\max_{k \leq K} \left| \frac{1}{1 + D_{1,k} \pi_{j,k}^N} - \frac{1}{1 + D_{j,k} \pi_j^N} \right| \to 0 \tag{40}$$

$$\max_{k \leq K} \left| \frac{D_{j,k}}{1 + D_{j,k} \pi_{j,k}^N} - \frac{D_{j,k}}{1 + D_{j,k} \pi_j^N} \right| \to 0 \tag{41}$$

$$\max_{k \leq K} \left| \frac{\pi_{j,k}^N}{1 + D_{j,k} \pi_{j,k}^N} - \frac{\pi_j^N}{1 + D_{j,k} \pi_j^N} \right| \to 0 \tag{42}$$

for $j = 1, 2$.

It now follows from (1), (38), (39), and (40)–(42) that $\left|\acute{G}_B^N(z,\alpha) - \acute{G}_B(z,\alpha)\right| \to 0$, $\left|\pi_1^N - \pi_1\right| \to 0$, and $\left|\pi_2^N - \pi_2\right| \to 0$, where $\acute{G}_B(z,\alpha)$, $\pi_1$, and $\pi_2$ satisfy

$$1 + z\acute{G}_B(z,\alpha) = \alpha\left(1 - \mathbf{E}\left[\frac{1}{1+\pi_j X_j}\right]\right) \quad , \; j = 1, 2. \tag{43}$$

$$\acute{G}_B(z,\alpha) = \frac{1}{z(z\pi_1\pi_2 - 1)} \tag{44}$$

To obtain the final solution we take $\alpha \to 1^-$ in (43) and (44) to obtain three simultaneous equations in the variables $G_B(z)$, $\pi_1$, and $\pi_2$, given by

$$G_B(z) = -z^{-1}\mathbf{E}\left[\frac{1}{1+\pi_j X_j}\right] \quad , \; j = 1, 2. \tag{45}$$

$$G_B(z) = \frac{1}{z(z\pi_1\pi_2 - 1)} \tag{46}$$

Since there exists a unique solution to these equations [22, Theorem 2.4], we have that $G_B^N(z) \to G_B(z)$ with probability 1.

It can easily be verified that this result matches that obtained using the S-transform from free probability theory [7]. Again, we emphasize that this derivation does not explicitly rely on free probability results; rather it relies on Assumptions 1–2.

## III. DS/MC-CDMA IN FREQUENCY-SELECTIVE CHANNELS

We now extend the incremental matrix expansion approach to determine the a.e.d. of a sum of matrices which are *not* free, corresponding to the received correlation matrix in a DS/MC-CDMA system. In doing so, we determine the asymptotic SINR of the MMSE receiver for this system.

### A. System Model

The following multi-user DS/MC-CDMA system model accounts for frequency-selective channels, and applies to

- the uplink of a single-cell system with multiple signatures per user[5] , or
- the downlink of a multi-cell system with a single (or multiple) signature(s) per user .

The received signal is given by

$$\mathbf{r} = \sum_{j=1}^{J} \mathbf{H}_j \mathbf{S}_j \mathbf{A}_j \mathbf{b}_j + \mathbf{n} \tag{47}$$

---

[5]This model with $\mathbf{A}_j = \mathbf{I}_{K_j}$ is considered in [18], however an approximation is used there to compute the SINR associated with isometric signatures.

where

- $\mathbf{H}_j$ is an $N \times N$ complex-valued matrix representing the channel from the $j^{\text{th}}$ transmitter to the base station. We assume that the matrices $\mathbf{H}_j \mathbf{H}_j^\dagger$, $j = 1, \ldots, J$ are *jointly diagonalizable*, that is, there exists a unitary $N \times N$ matrix $\mathbf{V}$ for which $\mathbf{V} \mathbf{H}_j \mathbf{H}_j^\dagger \mathbf{V}^\dagger$ is diagonal for all $j$. For MC-CDMA, each $\mathbf{H}_j$ is diagonal, and for DS-CDMA (with a cyclic prefix for each symbol) each $\mathbf{H}_j \mathbf{H}_j^\dagger$ is diagonalized by the discrete Fourier transform matrix. Assume $H_{\max} = \sup_N \max_{j \leq J} \|\mathbf{H}_j \mathbf{H}_j^\dagger\| < \infty$.

- $\mathbf{S}_j = [\mathbf{s}_{j,1} \cdots \mathbf{s}_{j,K_j}]$ is an $N \times K_j$ complex-valued signature-sequence matrix which contains either

    - random orthonormal columns, i.e., we assume that each $\mathbf{S}_j$ is obtained by extracting $K_j \leq N$ columns from an independent $N \times N$ Haar-distributed unitary random matrix, or,
    - i.i.d. complex elements[6] with mean zero and variance $\frac{1}{N}$ (for example, i.i.d. Gaussian real & imaginary parts $\sim \mathrm{N}(0, \frac{1}{2N})$), such that $\mathbf{S}_j$ is *unitarily invariant*.

  We shall call the first case 'isometric $\mathbf{S}_j$', and the second case 'i.i.d. $\mathbf{S}_j$', as in [12, 14]. Note that a *mixture* of i.i.d. and isometric signatures across $j$ is permitted in this model.

- $\mathbf{A}_j$ is a $K_j \times K_j$, diagonal, complex-valued matrix of transmit coefficients, i.e. $\mathbf{A}_j = \mathrm{diag}(A_{j,1}, \ldots, A_{j,K_j})$. In fact, the results which follow depend only on the values of $P_{j,k} = |A_{j,k}|^2$, and so to simplify notation, without lack of generality, we may assume $A_{j,k}$, $k = 1, \ldots, K$, is non-negative and real valued. Note that $P_{j,k}$ is the transmit power of the $k^{\text{th}}$ signature of transmitter $j$.

- The complex $K_j \times 1$ vector $\mathbf{b}_j$ contains the transmitted data symbols. Elements of $\mathbf{b}_m$ are assumed to be i.i.d. with zero mean and unit variance.

- $\mathbf{n}$ contains i.i.d., zero mean, circularly symmetric, complex Gaussian entries with variance per dimension $\sigma_n^2/2$.

- $\mathbf{H}_j$, $\mathbf{S}_j$, $\mathbf{A}_j$, $\mathbf{b}_j$, $j = 1, \ldots, J$ and $\mathbf{n}$ are mutually independent.

The output of the MMSE receiver for the $k^{\text{th}}$ signature of the $j^{\text{th}}$ transmitter is given by

$$\hat{\mathbf{b}}_j(k) = \mathbf{c}_{j,k}^\dagger \mathbf{r} \quad (48)$$

---

[6] For technical reasons, we also require that the elements have finite positive moments.

where

$$\mathbf{c}_{j,k} = A_{j,k}\mathbf{R}^{-1}\mathbf{H}_j\mathbf{s}_{j,k} \tag{49}$$

$$\mathbf{R} = \sigma_n^2 \mathbf{I}_N + \sum_{j=1}^{J} (\mathbf{H}_j \mathbf{S}_j \mathbf{A}_j)^{\ddagger} \tag{50}$$

Identifying the signal and interference components of the received signal in (47), i.e., $\mathbf{r} = A_{j,k}\mathbf{H}_j\mathbf{s}_{j,k}\mathbf{b}_j(k) + \mathbf{r}_I$, the corresponding output SINR is

$$\text{SINR}_{j,k}^N = \frac{\mathbf{E}[|\mathbf{c}_{j,k}^{\dagger}(\mathbf{r} - \mathbf{r}_I)|^2]}{\mathbf{E}[|\mathbf{c}_{j,k}^{\dagger}\mathbf{r}_I|^2]} \tag{51}$$

$$= P_{j,k}\rho_{j,k}^N \tag{52}$$

where the expectation in (52) is with respect to $\mathbf{n}$ and $\mathbf{b}_i$, $i = 1, \ldots, J$, and

$$\rho_{j,k}^N = \mathbf{s}_{j,k}^{\dagger}\mathbf{H}_j^{\dagger}\mathbf{R}_{d_{j,k}}^{-1}\mathbf{H}_j\mathbf{s}_{j,k} \tag{53}$$

where $\mathbf{R}_{d_{j,k}} = \mathbf{R} - (A_{j,k}\mathbf{H}_j\mathbf{s}_{j,k})^{\ddagger}$.

## B. Asymptotic MMSE SINR

We wish to evaluate the limiting SINR in (52) as $N$ and $K_j \to \infty$ with $K_j/N \to \alpha_j$ for each $j = 1, \ldots, J$. Under this limit, it is shown in Appendix IV that

$$\max_{k \leq K_j} |\rho_{j,k}^N - \rho_j^N| \xrightarrow{a.s.} 0 \tag{54}$$

where

$$\rho_j^N = \begin{cases} \frac{1}{N}\text{tr}[\mathbf{H}_j^{\dagger}\mathbf{R}^{-1}\mathbf{H}_j] & \text{, i.i.d. } \mathbf{S}_j, \\ \frac{1}{N-K_j}\text{tr}[\mathbf{\Upsilon}_j\mathbf{H}_j^{\dagger}\mathbf{R}^{-1}\mathbf{H}_j] & \text{, iso. } \mathbf{S}_j, \end{cases} \tag{55}$$

and $\mathbf{\Upsilon}_j = \mathbf{I}_N - \mathbf{S}_j\mathbf{S}_j^{\dagger}$.

Computing the limit of $\rho_j^N$ for $J = 1$ is well known [12, 23], and has been derived using an incremental matrix expansion approach in [14]. However, the extension to $J > 1$ is nontrivial. For $J > 1$ and $A_j = \mathbf{I}_{K_j}$, $j = 1, \ldots, J$, this problem is considered in [18], where the solution for i.i.d. signatures is obtained using [20, Theorem 16.3], and an approximate solution is derived for isometric signatures.

We now present an exact expression for the asymptotic SINR for $J > 1$ by extending the incremental matrix expansion approach. The following theorem is in terms of the Stieltjés transform of the e.d.f. of the eigenvalues of $\sum_{j=1}^{J}(\mathbf{H}_j\mathbf{S}_j\mathbf{A}_j)^{\ddagger}$, from which the asymptotic MMSE SINR is an auxiliary result. That is, we redefine $\mathbf{R} = -z\mathbf{I}_M + \sum_{j=1}^{J}(\mathbf{H}_j\mathbf{S}_j\mathbf{A}_j)^{\ddagger}$, where $z \in \mathbb{C}^+$, such that the Stieltjés transform of the e.d.f. of the eigenvalues of $\sum_{j=1}^{J}(\mathbf{H}_j\mathbf{S}_j\mathbf{A}_j)^{\ddagger}$ is given by $G_{\mathsf{R}}^N(z) = \frac{1}{M}\text{tr}[\mathbf{R}^{-1}]$.

The theorem is given in terms of the $2J$ additional random variables $\rho_j^N \in \mathbb{C}^+$, $j = 1 \ldots J$, as defined in (55), using the redefinition of $\mathbf{R}$ mentioned above, and $\tau_j^N \in \mathbb{C}^+$, $j = 1 \ldots J$. The variable $\tau_j^N$ is defined in terms of matrix equations in Appendix IV, however, as the definition of $\tau_j^N$ is lengthy, and is not needed to state the following result, to facilitate the flow of results it is not stated here.

**Theorem 1** *Assume that the e.d.f.s of the eigenvalues of $\mathbf{H}_j\mathbf{H}_j^{\dagger}$ and $\mathbf{A}\mathbf{A}^{\dagger}$ converge in distribution almost surely to compactly supported probability measures on $\mathbb{R}^*$ as $(N, K_j) \to \infty$ with $K_j/N \to \alpha_j > 0$, $j = 1, \ldots, J$. Then the Stieltjés transform of the e.d.f. of the eigenvalues of $\sum_{j=1}^{J}(\mathbf{H}_j\mathbf{S}_j\mathbf{A}_j)^{\ddagger}$, $G_R^N(z)$, $z \in \mathbb{C}^+$, along with $\rho_j^N$ and $\tau_j^N$, $j = 1 \ldots J$ satisfy*

$$\left|G_R^N(z) - G_R(z)\right| \xrightarrow{a.s.} 0, \tag{56}$$

$$\left|\rho_j^N - \rho_j\right| \xrightarrow{a.s.} 0, \qquad j = 1 \ldots J, \tag{57}$$

$$\left|\tau_j^N - \tau_j\right| \xrightarrow{a.s.} 0, \qquad j = 1 \ldots J, \tag{58}$$

*where $G_R(z), \rho_j, \tau_j \in \mathbb{C}^+$ are solutions to*

$$G_R(z) = -\frac{1}{z}\left(1 - \sum_{j=1}^{J}\alpha_j\rho_j\mathcal{P}_j\right) \tag{59}$$

$$\rho_j = \begin{cases} \mathcal{H}_j & , \text{ i.i.d. } \mathbf{S}_j, \\ \dfrac{\mathcal{H}_j}{1 - \alpha_j\rho_j\mathcal{P}_j} & , \text{ iso. } \mathbf{S}_j. \end{cases} \tag{60}$$

$$\tau_j = \begin{cases} \alpha_j(\bar{p}_j - \mathcal{P}_j) & , \text{ i.i.d. } \mathbf{S}_j, \\ \alpha_j(\bar{p}_j - \mathcal{P}_j) - (\alpha_j\bar{p}_j - \tau_j)^2\mathcal{H}_j & , \text{ iso. } \mathbf{S}_j. \end{cases} \tag{61}$$

*where*

$$\mathcal{P}_j = \mathsf{E}\left[\frac{P_j}{1 + P_j\rho_j}\right] \tag{62}$$

$$\mathcal{H}_j = \mathsf{E}\left[\frac{H_j}{-z + \sum_i(\alpha_i\bar{p}_i - \tau_i)H_i}\right] \tag{63}$$

*and the expectation in (63) is with respect to $\{H_i\}_{i=1,\ldots,J}$, where $H_i$ is a scalar random variable according to the a.e.d. of $\mathbf{H}_i\mathbf{H}_i^{\dagger}$. Similarly, the expectation in (62) is with respect to $P_j$, a scalar random variable according to the a.e.d. of $\mathbf{A}_j\mathbf{A}_j^{\dagger}$, and $\bar{p}_j = \mathsf{E}[P_j]$.*

*Proof:* See Appendix IV. ∎

We have the following remarks concerning Theorem 1:

- If (59)–(63) has a unique solution $G_R(z), \rho_j, \tau_j \in \mathbb{C}^+$ for any given $z \in \mathbb{C}^+$, then Theorem 1 additionally gives that the e.d.f. of the eigenvalues of $(\mathbf{HSA})^\ddagger$ almost surely converges in distribution to a deterministic distribution, whose Stieltjés transform is $G_R(z)$. Moreover, we have that $\rho_j^N$ converges almost surely to the deterministic value $\rho_j$ in the limit considered, and so, letting $z = -\sigma_n^2 + \epsilon j$ and taking $\epsilon \to 0$, as indicated by (52), (54), and (57), the asymptotic SINR of the $k^{\text{th}}$ data stream of the $j^{\text{th}}$ transmitter almost surely converges to $P_{j,k}\rho_j$.

- The solution, at a given value of $z$, requires solving the $2J+1$ equations (59),(60), and (61), which contain $2J+1$ variables, i.e., $G_R(z)$, $\rho_j$, and $\tau_j$ for $j = 1, \ldots, J$.

- For the special case of $\mathbf{A}_j = \mathbf{I}_{K_j}$ and i.i.d. $\mathbf{S}_j$ for all $j = 1, \ldots, J$, Theorem 1 can also be obtained from [20, Theorem 16.3].[7]

- To find the SINR, the approximate method in [18] requires solving $J$ sets of equations, each of which contains $2J+1$ variables. In contrast, Theorem 1 states that solving one set of equations in $J$ (independent) variables gives the SINRs for all signatures of all transmitters.

- If the channels of the transmitters are independent, then the expectation in (63) becomes

$$\mathcal{H}_j = \int \cdots \int \frac{h_j}{-z + \sum_i(\alpha_i - \tau_i)h_i} dF_{H_1}(h_1) \ldots dF_{H_J}(h_J) \tag{64}$$

If we further assume that the a.e.d. of each $\mathbf{H}_j\mathbf{H}_j^\dagger$ is discrete, i.e., has the form

$$f_{H_m}(h_m) = \sum_{n=1}^{N_p} \beta_{m,n}\delta(h_m - p_{m,n}) \tag{65}$$

where $\beta_{m,n} \in [0,1]$ and $\sum_{n=1}^{N_p} \beta_{m,n} = 1$ for each $m = 1, \ldots, J$, and $N_p$ is some finite positive integer, then

$$\mathcal{H}_j = \sum_{n_1=1}^{N_p} \cdots \sum_{n_J=1}^{N_p} \frac{p_{j,n_j} \prod_{i=1}^J \beta_{i,n_i}}{-z + \sum_{i=1}^J(\alpha_i - \tau_i)p_{i,n_i}} \tag{66}$$

## C. Example

Consider two equal-power transmitters with $\alpha_1 = \alpha_2$, where $H_j$ in (63) is exponentially distributed with unit mean. Figure 1 shows empirical values (numerically generated from averaging finite systems with $N = 32$ and QPSK modulation) and asymptotic values (determined from Theorem 1) of $G_R(z)$ and MMSE SINR. Also shown are the values obtained using the approximate asymptotic results of [18]. As expected, the exact solution matches the empirical values, and moreover, the approximate solution of [18] is seen to be very accurate. As mentioned previously, the computational complexity of the exact solution in Theorem 1 is significantly less than that of the approximate solution.

---

[7]The authors thank P. Loubaton for pointing this out.

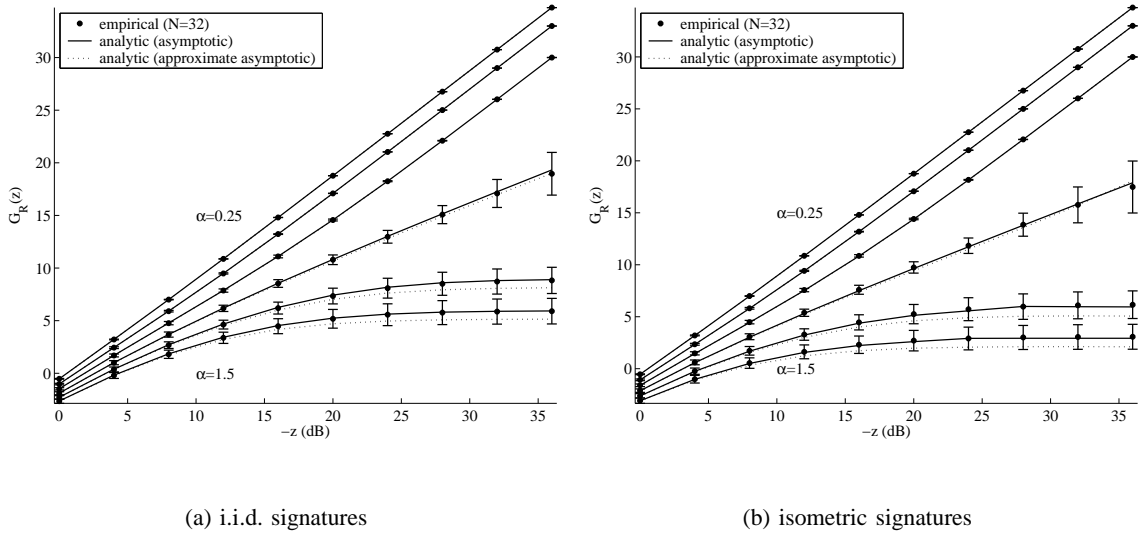

(a) i.i.d. signatures  (b) isometric signatures

Fig. 1. $G_{\mathbf{R}}(z)$ vs. $z$: Asymptotic and empirical ($N = 32$, BPSK, $5 \times 10^3$ realizations) values, for two transmitters, $\alpha_1 = \alpha_2$, $\mathbf{A}_1 = \mathbf{A}_2 = \mathbf{I}_{K/2}$, $\mathbf{E}[H_1] = \mathbf{E}[H_2]$, $\text{Im}(z) \to 0$, for $\alpha = 0.25$ to $1.5$ in steps of $0.25$.

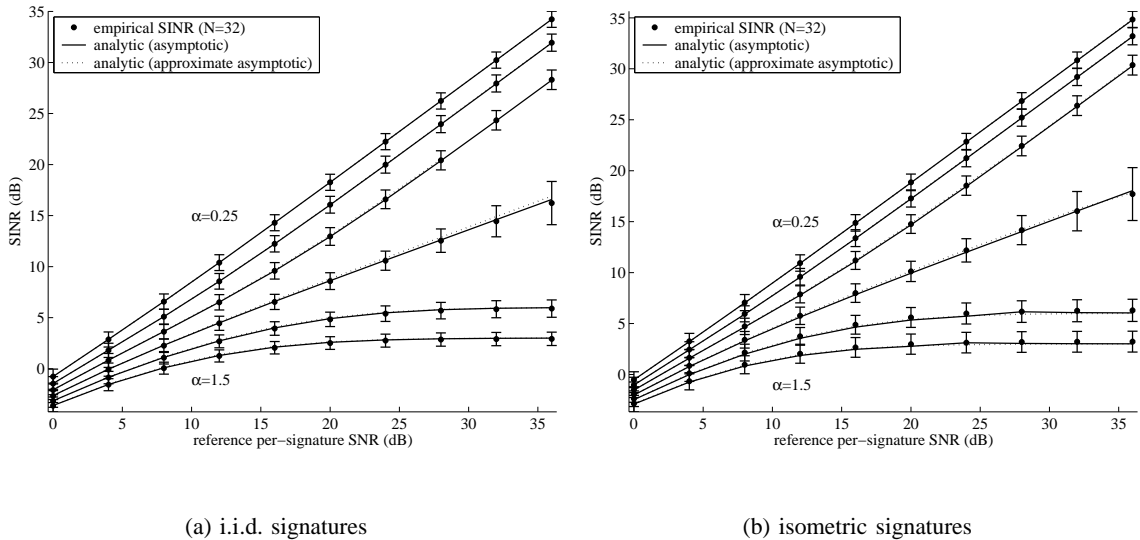

(a) i.i.d. signatures  (b) isometric signatures

Fig. 2. SINR vs. SNR: Same parameters as for Figure 1.

## IV. CONCLUSIONS

Using the approach of [14], we have evaluated the a.e.d.'s of sums and products of unitarily invariant matrices, and obtained the same result as given by free probability theory. The derivation given here is significantly simpler than the general proof for free non-commuting random variables as derived by Voiculescu, and indicates that for random matrices, "free" is synonymous with "unitarily invariant".

We also derived the asymptotic distribution of a sum of non-free random matrices, corresponding to the receive autocorrelation matrix for a class of DS/MC-CDMA systems. This result is primarily due to an extension of [14, Lemma 6], given in Lemma 12, which is interesting in its own right. Interestingly, the exact asymptotic results are very close to the results obtained if the non-free component matrices are (incorrectly) assumed to be free, although the accuracy of this approximation is not well understood. The applicability of this approach to other channel models (such as dispersive and possibly correlated multi-user MIMO channels, etc.) remains to be determined. In particular, the current approach requires $\mathbf{H}_j\mathbf{H}_j^\dagger$, $j = 1, \ldots, J$, to be jointly diagonalizable.

## V. ACKNOWLEDGEMENTS

The authors thank P. Loubaton for bringing references [20, 22] to our attention.

## APPENDIX I

## AUXILIARY RESULTS

For the derivations which follow, we recall the following definitions and results from [14, Appendix I] concerning *asymptotic equivalence* and *uniform asymptotic equivalence* of random sequences.

### A. Asymptotic Equivalence

**Definition 1** *Let $\{a_N\}_{N=1,\ldots}$ and $\{b_N\}_{N=1,\ldots}$ denote a pair of infinite sequences of complex-valued random variables indexed by $N$. These sequences are defined to be* asymptotically equivalent, *denoted $a_N \asymp b_N$, iff $|a_N - b_N| \xrightarrow{a.s.} 0$ as $N \to \infty$, where $\xrightarrow{a.s.}$ denotes almost-sure convergence in the limit considered.* □

Clearly $\asymp$ is an equivalence relation, transitivity being obtained through the triangle inequality. We shall additionally define asymptotic equivalence for sequences of $N \times 1$ vectors and $N \times N$ matrices in an identical manner as above, where the absolute value is replaced by the Euclidean vector norm and the associated induced spectral norm, respectively.

**Lemma 1** *If $a_N \asymp b_N$ and $x_N \asymp y_N$, and if $|a_N|$, $|y_N|$ and/or $|b_N|$, $|x_N|$ are almost surely uniformly bounded above[8] over $N$, then $a_N x_N \asymp b_N y_N$. Similarly, $a_N/x_N \asymp b_N/y_N$ if $|a_N|$ or $|b_N|$ is uniformly bounded above over $N$, and at least one of $\inf_N |x_N|$ and $\inf_N |y_N|$ is positive almost surely.*

Note that the multiplicative part of Lemma 1 holds for any mixture of matrices, vectors or scalars for which the dimensions of $a_N$ and $x_N$ are such that $a_N x_N$ makes sense, due to the submultiplicative property of the spectral norm. The following definition and related results, however, are concerned with scalar complex sequences.

**Definition 2** *Let $\{\{a_{N,n}\}_{n=1...N}\}_{N=1,...}$ and $\{\{b_{N,n}\}_{n=1...N}\}_{N=1,...}$ denote a pair of infinite sequences, indexed by $N$. The $N^{th}$ element is a complex-valued sequences of length $N$, indexed by $n$. These sequences are defined to be* uniformly asymptotically equivalent, *denoted $a_{N,n} \stackrel{n}{\asymp} b_{N,n}$, iff $\max_{n \leq N} |a_{N,n} - b_{N,n}| \stackrel{a.s.}{\longrightarrow} 0$ as $N \to \infty$.* □

Also, we define $a_N$ and $b_{N,n}$ as being uniformly asymptotically equivalent (denoted $a_N \stackrel{n}{\asymp} b_{N,n}$), if $a_{N,n} \stackrel{n}{\asymp} b_{N,n}$ where $a_{N,n} = a_N$ for all $n = 1, \ldots, N$.

Also, analogous to Lemma 1, we have

**Lemma 2** *If $a_{N,n} \stackrel{n}{\asymp} b_{N,n}$ and $x_{N,n} \stackrel{n}{\asymp} y_{N,n}$, and if $|a_{N,n}|$, $|y_{N,n}|$ and/or $|b_{N,n}|$, $|x_{N,n}|$ are almost surely uniformly bounded above over $N$ and $n$, then $a_{N,n} x_{N,n} \stackrel{n}{\asymp} b_{N,n} y_{N,n}$. Similarly, $a_{N,n}/x_{N,n} \stackrel{n}{\asymp} b_{N,n}/y_{N,n}$ if $|a_{N,n}|$ or $|b_{N,n}|$ is almost surely uniformly bounded above over $N$ and $n$, and at least one of $\inf_{N,n} |x_{N,n}|$ and $\inf_{N,n} |y_{N,n}|$ is positive almost surely.*

*B. Proving Asymptotic Equivalence*

The following results are required in order to prove asymptotic equivalence.

**Theorem 2** *If $\mathsf{E}[|X|^r] < \infty$ for $r > 0$ (not necessarily an integer); then*

$$P_X(|X| \geq \epsilon) \leq \frac{\mathsf{E}[|X|^r]}{\epsilon^r} \tag{67}$$

*for every $\epsilon > 0$.*

For $r = 1$, Theorem 2 is often called Markov's inequality. For $X$ substituted by $(X - \mathsf{E}[X])^2$, where $X$ has finite mean and variance, and $r = 1$, Theorem 2 is often called Chebyshev's inequality.

---

[8]A sequence $\{a_N\}_{N=1...}$ of complex-valued $N \times 1$ vectors or scalars is uniformly bounded above over $N$ if $\sup_N |a_N| < \infty$, or in the case of complex-valued $N \times N$ matrices, $\sup_N \|a_N\| < \infty$.

**Lemma 3 (The Borel-Cantelli lemma)** *Let $\{E_N \in \mathcal{F}\}_{N=1,2,\ldots}$ denote a sequence of events in the probability space $(\Omega, \mathcal{F}, \mathcal{P})$. If*

$$\sum_{N=1}^{\infty} \mathcal{P}(E_N) < \infty \tag{68}$$

*then the probability that an infinite number of the $E_i$'s occur is zero.*

**Example 1** Showing almost-sure convergence of the sequence $\{X_N\}_{N=1,2\ldots}$ reduces to demonstrating that $\mathbf{E}[|X_N|^m]$ for some $m \geq 1$ is $\mathrm{O}(N^{-n})$, where $n \geq 2$ (typically, $m = 4$ and $n = 2$). Then, from Theorem 2 we have

$$P_{X_N}(|X_N| \geq \epsilon) \leq \frac{c}{\epsilon^r N^n} \tag{69}$$

for $\epsilon > 0$ and some finite, positive $c$, independent of $N$. Moreover, (69) implies

$$\sum_{N=1}^{\infty} P_{X_N}(|X_N| \geq \epsilon) < \infty \tag{70}$$

due to the fact that $n \geq 2$. Finally, from (70) and Lemma 3, $X_N$ converges to zero almost surely as $N \to \infty$.

**Lemma 4** [8, Lemma 1] *Let $\mathbf{C}_N$, be an $N \times N$ complex-valued matrix with uniformly bounded spectral radius for all $N$, i.e., $\sup_N \|\mathbf{C}_N\| < \infty$, and $\mathbf{y} = [X_1, \ldots, X_N]^\dagger / \sqrt{N}$, where the $X_i$'s are i.i.d. complex random variables with mean zero, unit variance, and finite eighth moment. Then*

$$\mathbf{E}[|\mathbf{y}^\dagger \mathbf{C} \mathbf{y} - \mathrm{tr}[\mathbf{C}]|^4] \leq \frac{c}{N^2} \tag{71}$$

*where the constant $c > 0$ does not depend on $N$, $\mathbf{C}$, nor on the distribution of the $X_i$.*

**Lemma 5** *Let $\mathbf{S}$ be $K < N$ columns of an $N \times N$ Haar distributed random matrix, and suppose $\mathbf{s}$ is a column of $\mathbf{S}$. Let $\mathbf{X}_N$ be an $N \times N$ random matrix, which is a non-trivial function of all columns of $\mathbf{S}$ except $\mathbf{s}$, and $B = \sup_N \|\mathbf{X}_N\| < \infty$. Then,*

$$\mathbf{E}\left[\left|\mathbf{s}^\dagger \mathbf{X}_N \mathbf{s} - \frac{1}{N-K} \mathrm{tr}[\mathbf{\Pi} \mathbf{X}_N]\right|^4\right] \leq \frac{C}{N^2} \tag{72}$$

*where $\mathbf{\Pi} = \mathbf{I}_N - (\mathbf{S}\mathbf{S}^\dagger - \mathbf{s}\mathbf{s}^\dagger)$ and $C$ is a deterministic finite constant which depends only on $B$ and $\alpha = K/N$.*

   *Proof:* This result is a straightforward extension of [13, Proposition 4].  ∎

Due to Theorem 2 and Lemma 3, an immediate consequence of Lemmas 4 and 5 is that $\mathbf{y}^\dagger \mathbf{C} \mathbf{y} \asymp \mathrm{tr}[\mathbf{C}]$ and $\mathbf{s}^\dagger \mathbf{X}_N \mathbf{s} \asymp \frac{1}{N-K} \mathrm{tr}[\mathbf{\Pi} \mathbf{X}_N]$ respectively, as explained in Example 1 above.

**Lemma 6** *Suppose $a_{N,n} \overset{n}{\asymp} b_N$ as defined in Definition 2. Let $\{\{c_{N,n}\}_{n=1...N}\}_{N=1,...}$ denote an infinite sequence, indexed by $N$, of sets of* real-valued *sequences of length $N$, indexed by $n$. Additionally assume that*

$$\delta = \inf_{N,n} |\text{Im}(a_{N,n})| > 0 \tag{73}$$

*almost surely, and that $a_{N,n}$ and/or $b_N$ is uniformly bounded above over $N$ and $n$. Then*

$$\frac{1}{1+c_{N,n}a_{N,n}} \overset{n}{\asymp} \frac{1}{1+c_{N,n}b_N} \tag{74}$$

$$\frac{a_{N,n}}{1+c_{N,n}a_{N,n}} \overset{n}{\asymp} \frac{b_N}{1+c_{N,n}b_N} \tag{75}$$

$$\frac{c_{N,n}}{1+c_{N,n}a_{N,n}} \overset{n}{\asymp} \frac{c_{N,n}}{1+c_{N,n}b_N} \tag{76}$$

*Proof:* Firstly, if $c_{N,n} = 0$ for any $N$ and $n \leq N$, then (74) is clearly true. So, consider $c_{N,n} \neq 0$. Denote $B = \sup_N |b_N| < \infty$. Consider a realization for which $\max_{n \leq N} |a_{N,n} - b_N| \to 0$ holds. Note that for any $N$ and $n \leq N$, $\delta \leq |\text{Im}(a_{N,n})| \leq |\text{Im}(a_{N,n} - b_N)| + |\text{Im}(b_N)|$. Hence, take $N$ sufficiently large such that $\max_{n \leq N} |\text{Im}(a_{N,n} - b_N)| \leq \delta/2$, so that $|\text{Im}(b_N)| \geq \delta/2$. Moreover, note that $|1 + c_{N,n}b_N| \geq |c_{N,n} \text{Im}(b_N)| \geq |c_{N,n}|\delta/2$, and similarly $|1 + c_{N,n}a_{N,n}| \geq |c_{N,n}|\delta$, due to the inequality $|1 + Dx| \geq |D|\text{Im}(x)$ for $D \in \mathbb{R}$ and $x \in \mathbb{C}^+$. We then obtain

$$\left| \frac{1}{1+c_{N,n}a_{N,n}} - \frac{1}{1+c_{N,n}b_N} \right| = \left| \frac{c_{N,n}a_{N,n}}{1+c_{N,n}a_{N,n}} - \frac{c_{N,n}b_N}{1+c_{N,n}b_N} \right|$$

$$\leq \frac{|c_{N,n}||a_{N,n} - b_N|}{|1+c_{N,n}a_{N,n}|} + \frac{|c_{N,n}|^2 |b_N||a_{N,n} - b_N|}{|1+c_{N,n}a_{N,n}||1+c_{N,n}b_N|}$$

$$\leq \frac{1}{\delta}\left(1 + \frac{2B}{\delta}\right) |a_{N,n} - b_N| \tag{77}$$

Taking a maximum over $n$, and using the facts that $a_{N,n} \overset{n}{\asymp} b_N$, $B < \infty$, and $\delta > 0$ gives (74), assuming $b_N$ is uniformly bounded above for all $N$. The remaining case, where $a_{N,n}$ is uniformly bounded above, is shown in an identical manner.

To show (75), note that

$$\frac{1}{|1+c_{N,n}a_{N,n}|} \leq 1 + \frac{|c_{N,n}a_{N,n}|}{|1+c_{N,n}a_{N,n}|} \leq 1 + \frac{2B}{\delta}. \tag{78}$$

Using (77) and (78) gives

$$\left| \frac{a_{N,n}}{1+c_{N,n}a_{N,n}} - \frac{b_N}{1+c_{N,n}b_N} \right| \leq \frac{|a_{N,n} - b_N|}{|1+c_{N,n}a_{N,n}|} + |b_N| \left| \frac{1}{1+c_{N,n}a_{N,n}} - \frac{1}{1+c_{N,n}b_N} \right| \tag{79}$$

$$\leq \left(1 + \frac{2B}{\delta}\right)\left(1 + \frac{B}{\delta}\right) |a_{N,n} - b_N| \tag{80}$$

which implies (75). Finally, (76) is obtained due to

$$\left| \frac{c_{N,n}}{1+c_{N,n}a_{N,n}} - \frac{c_N}{1+c_{N,n}b_N} \right| \leq \frac{2}{\delta^2} |a_{N,n} - b_N| \tag{81}$$

∎

**Lemma 7** *If $a_{N,n} \stackrel{n}{\asymp} b_{N,n}$ as defined in Definition 2, then $\frac{1}{N}\sum_{n=1}^{N} a_{N,n} \asymp \frac{1}{N}\sum_{n=1}^{N} b_{N,n}$.*

*Proof:* This follows immediately from

$$\left| \frac{1}{N}\sum_{n=1}^{N}(a_{N,n} - b_{N,n}) \right| \leq \max_{n \leq N} |a_{N,n} - b_{N,n}| \tag{82}$$

∎

**Lemma 8** [14, Lemma 5] *For $N = 1,\ldots$, let $\mathbf{X}_N = \mathbf{M}_N - z\mathbf{I}_N$, where $\mathbf{M}_N$ is an $N \times N$ Hermitian matrix and $z \in \mathbb{C}^+$, and suppose $\mathbf{u}_N \in \mathbb{C}^N$. Denote $u_N = \mathbf{u}_N^\dagger \mathbf{X}_N^{-1} \mathbf{u}_N$. If*

$$b = \inf_N |\mathbf{u}_N| > 0 \quad a.s. \tag{83}$$

$$B = \sup_N \|\mathbf{X}_N\| < \infty \quad a.s. \tag{84}$$

*Then*

$$\mathrm{Im}(u_N) \geq \mathrm{Im}(z) \frac{b^2}{B^2} \quad a.s. \tag{85}$$

*and hence $u_N \in \mathbb{C}^+$, almost surely.*

**Lemma 9** [2, Lemma 2.6] *Let $z \in \mathbb{C}^+$, $\mathbf{A}$ and $\mathbf{B}$ $N \times N$ Hermitian, $\tau \in \mathbb{R}$, and $\mathbf{q} \in \mathbb{C}^N$. Then,*

$$\left| \mathrm{tr}\left[ \left( (\mathbf{B} - z\mathbf{I})^{-1} - (\mathbf{B} + \tau\mathbf{q}\mathbf{q}^\dagger - z\mathbf{I})^{-1} \right) \mathbf{A} \right] \right| \leq \frac{\|\mathbf{A}\|}{\mathrm{Im}(z)}. \tag{86}$$

*C. Asymptotic extensions of the matrix inversion lemma*

**Lemma 10** [14, Lemma 6] *Let $\mathbf{Y}_N = \mathbf{X}_N + \mathbf{v}_N\mathbf{u}_N^\dagger + \mathbf{u}_N\mathbf{v}_N^\dagger + c_N\mathbf{u}_N\mathbf{u}_N^\dagger$, where $\mathbf{v}_N, \mathbf{u}_N \in \mathbb{C}^N$, $c_N \in \mathbb{R}^*$, and $\mathbf{X}_N = \mathbf{M}_N - z\mathbf{I}_N$, where $\mathbf{M}_N$ is an $N \times N$ Hermitian matrix and $z \in \mathbb{C}^+$. Denote*

$$\epsilon_N = \mathbf{u}_N^\dagger \mathbf{X}_N^{-1} \mathbf{v}_N \tag{87}$$

$$u_N = \mathbf{u}_N^\dagger \mathbf{X}_N^{-1} \mathbf{u}_N \tag{88}$$

$$v_N = \mathbf{v}_N^\dagger \mathbf{X}_N^{-1} \mathbf{v}_N \tag{89}$$

*Assume that as $N \to \infty$,*

$$|\epsilon_N| \xrightarrow{a.s.} 0 \tag{90}$$

*and*

$$b = \inf_N |\mathbf{u}_N| > 0 \quad , a.s., \tag{91}$$

$$B = \sup_N \max\{\|\mathbf{X}_N\|, |\mathbf{v}_N|, |\mathbf{u}_N|, |c_N|\} < \infty \tag{92}$$

*Then,*

$$\left| \mathbf{Y}_N^{-1}\mathbf{u}_N - \frac{\mathbf{X}_N^{-1}(\mathbf{u}_N - u_N\mathbf{v}_N)}{1 - u_N(v_N - c_N)} \right| \xrightarrow{a.s.} 0 \tag{93}$$

$$\left| \mathbf{Y}_N^{-1}\mathbf{v}_N - \frac{\mathbf{X}_N^{-1}(-v_N\mathbf{u}_N + (1 + c_N u_N)\mathbf{v}_N)}{1 - u_N(v_N - c_N)} \right| \xrightarrow{a.s.} 0 \tag{94}$$

*as $N \to \infty$, and*

$$\delta = \inf_N |1 - u_N(v_N - c_N)| > 0 \tag{95}$$

*almost surely, where $\delta$ depends only on B, b, and $\mathrm{Im}(z)$.*

**Lemma 11** [14, Lemma 7] *Let $\mathbf{A}_N$ be an $N \times N$ Hermitian matrix, and suppose $A = \sup_N \|\mathbf{A}_N\| < \infty$. Using the definitions and assumptions of Lemma 10, additionally define*

$$\varepsilon_N^{(1)} = \mathbf{u}_N^\dagger \mathbf{X}_N^{-1} \mathbf{A}_N \mathbf{X}_N^{-1} \mathbf{v}_N \tag{96}$$

$$\varepsilon_N^{(2)} = \mathbf{v}_N^\dagger \mathbf{X}_N^{-1} \mathbf{A}_N \mathbf{X}_N^{-1} \mathbf{u}_N \tag{97}$$

$$\acute{u}_N = \mathbf{u}_N^\dagger \mathbf{X}_N^{-1} \mathbf{A}_N \mathbf{X}_N^{-1} \mathbf{u}_N \tag{98}$$

$$\acute{v}_N = \mathbf{v}_N^\dagger \mathbf{X}_N^{-1} \mathbf{A}_N \mathbf{X}_N^{-1} \mathbf{v}_N \tag{99}$$

*Then,*

$$\left| \mathrm{tr}[\mathbf{A}_N \mathbf{Y}_N^{-1}] - \left( \mathrm{tr}[\mathbf{A}_N \mathbf{X}_N^{-1}] + \frac{u_N \acute{v}_N + (v_N - c_N)\acute{u}_N - \varepsilon_N^{(1)} - \varepsilon_N^{(2)}}{1 - u_N(v_N - c_N)} \right) \right| \xrightarrow{a.s.} 0 \tag{100}$$

*as $N \to \infty$.*

We require the following extension to Lemma 10.

**Lemma 12** *Let $\mathbf{Y}_N = \mathbf{X}_N + \sum_{j=1}^J \left( \mathbf{v}_{N,j}\mathbf{u}_N^\dagger + \mathbf{u}_N^\dagger \mathbf{v}_{N,j} + c_{N,j}\mathbf{u}_N \mathbf{u}_N^\dagger \right)$, where $\mathbf{u}_N, \mathbf{v}_{N,j} \in \mathbb{C}^N$, $c_{N,j} \in \mathbb{R}^*$, $J$ is a finite positive integer, and $\mathbf{X}_N = \mathbf{M}_N - z\mathbf{I}_N$, where $\mathbf{M}_N$ is an $N \times N$ Hermitian matrix and $z \in \mathbb{C}^+$. Denote*

$$\epsilon_{N,j} = \mathbf{u}_N^\dagger \mathbf{X}_N^{-1} \mathbf{v}_{N,j}, \tag{101}$$

$$u_N = \mathbf{u}_N^\dagger \mathbf{X}_N^{-1} \mathbf{u}_N, \tag{102}$$

$$v_{N,j} = \mathbf{v}_{N,j}^\dagger \mathbf{X}_N^{-1} \mathbf{v}_{N,j}, \tag{103}$$

$$\chi_{N,i,j} = \mathbf{v}_{N,i}^\dagger \mathbf{X}_N^{-1} \mathbf{v}_{N,j} \quad , i \neq j. \tag{104}$$

*Assume that as* $N \to \infty$,

$$|\epsilon_{N,j}| \xrightarrow{a.s.} 0, \tag{105}$$

$$|\chi_{N,i,j}| \xrightarrow{a.s.} 0 \quad , i \neq j, \tag{106}$$

*for all* $i, j \in \{1, \ldots, J\}$, *and*

$$\inf_N |u_N| > 0 \quad , a.s., \tag{107}$$

$$B = \sup_N \max_{j \leq J} \max \{\|\mathbf{X}_N\|, |u_N|, |\mathbf{v}_{N,j}|, |c_{N,j}|\} < \infty \tag{108}$$

*Let* $d_{N,j} = c_{N,j} - v_{N,j}$. *Then,*

$$\left| \mathbf{Y}_N^{-1} \mathbf{u}_N - \frac{\mathbf{X}^{-1}\left(\mathbf{u}_N - u_N \sum_j \mathbf{v}_{N,j}\right)}{1 + u_N \sum_j d_{N,j}} \right| \xrightarrow{a.s.} 0 \tag{109}$$

$$\left| \mathbf{Y}_N^{-1} \mathbf{v}_{N,i} - \frac{\mathbf{X}^{-1}\left(-v_{N,i}\mathbf{u}_N + \left(1 + u_N\left(c_{N,i} + \sum_{j \neq i} d_{N,j}\right)\right)\mathbf{v}_{N,i} + v_{N,i} u_N \sum_{j \neq i} \mathbf{v}_{N,j}\right)}{1 + u_N \sum_j d_{N,j}} \right| \xrightarrow{a.s.} 0 \tag{110}$$

*as* $N \to \infty$, *and*

$$\delta = \inf_N \left| 1 + u_N \sum_j d_{N,j} \right| > 0 \tag{111}$$

*almost surely, where* $\delta$ *depends only on* $B$, $b$, *and* $\text{Im}(z)$.

*Proof:* See Appendix V. ∎

**Lemma 13** *Let* $\mathbf{A}_N$ *be an* $N \times N$ *Hermitian matrix, and suppose* $A = \sup_N \|\mathbf{A}_N\| < \infty$. *In addition to the definitions and assumptions of Lemma 12, define*

$$\varepsilon_{N,j}^{(1)} = \mathbf{u}_N^\dagger \mathbf{X}_N^{-1} \mathbf{A}_N \mathbf{X}_N^{-1} \mathbf{v}_{N,j} \tag{112}$$

$$\varepsilon_{N,j}^{(2)} = \mathbf{v}_{N,j}^\dagger \mathbf{X}_N^{-1} \mathbf{A}_N \mathbf{X}_N^{-1} \mathbf{u}_N \tag{113}$$

$$\acute{u}_N = \mathbf{u}_N^\dagger \mathbf{X}_N^{-1} \mathbf{A}_N \mathbf{X}_N^{-1} \mathbf{u}_N \tag{114}$$

$$\acute{v}_{N,j} = \mathbf{v}_{N,j}^\dagger \mathbf{X}_N^{-1} \mathbf{A}_N \mathbf{X}_N^{-1} \mathbf{v}_{N,j} \tag{115}$$

$$\acute{\chi}_{N,i,j} = \mathbf{v}_{N,i}^\dagger \mathbf{X}_N^{-1} \mathbf{A}_N \mathbf{X}_N^{-1} \mathbf{v}_{N,j} \quad , i \neq j. \tag{116}$$

*and assume that as* $N \to \infty$

$$|\acute{\chi}_{N,i,j}| \xrightarrow{a.s.} 0 \quad , i \neq j \tag{117}$$

*for all* $i, j \in \{1, \ldots, J\}$.

*Then,*

$$\left| \text{tr}[\mathbf{A}_N \mathbf{Y}_N^{-1}] - \left( \text{tr}[\mathbf{A}_N \mathbf{X}_N^{-1}] + \frac{u_N \sum_j \acute{v}_{N,j} - \acute{u}_N \sum_j d_{N,j} - \sum_j (\varepsilon_{N,j}^{(1)} + \varepsilon_{N,j}^{(2)})}{1 + u_N \sum_j d_{N,j}} \right) \right| \xrightarrow{a.s.} 0 \quad (118)$$

*Proof:* This can be shown from Lemma 11 and Lemma 12 using induction on $J$. ∎

## APPENDIX II

### PROOFS FOR SECTION II-B.

We first show $\max_{k \leq K} \left| \rho_{j,k}^N - \rho_j^N \right| \xrightarrow{a.s.} 0$ in the limit considered. Define

$$\rho_{j,k}^{N'} = \frac{1}{N-K} \text{tr}[\mathbf{\Upsilon}_{j,k} \acute{\mathbf{C}}_{j,k}^{-1}] \quad (119)$$

$$\rho_{j,k}^{N''} = \frac{1}{N-K} \text{tr}[\mathbf{\Upsilon}_j \acute{\mathbf{C}}_{j,k}^{-1}] \quad (120)$$

$$\mathbf{\Upsilon}_{j,k} = \mathbf{\Upsilon}_j + \mathbf{v}_{j,k} \mathbf{v}_{j,k}^\dagger \quad (121)$$

From Lemma 5, Theorem 2, and Lemma 3, we have

$$\max_{k \leq K} \left| \rho_{j,k}^N - \rho_{j,k}^{N'} \right| \xrightarrow{a.s.} 0 \quad (122)$$

in the limit considered, as explained in Example 1 in Appendix I. We now consider a realization for which (122) holds. We obtain

$$\left| \rho_{j,k}^{N'} - \rho_{j,k}^{N''} \right| = \frac{1}{N-K} |\rho_{j,k}^N| \leq \frac{2}{N-K} \left| \rho_{j,k}^{N'} \right| \quad (123)$$

$$\leq \frac{2}{(N-K)^2} \| \mathbf{\Upsilon}_{j,k} \acute{\mathbf{C}}_{j,k}^{-1} \| \, \text{rank}\,(\mathbf{\Upsilon}_{j,k}) \quad (124)$$

$$< \frac{2}{\text{Im}(z)(N-K)} \quad (125)$$

where the first inequality follows from (122) with $N$ sufficiently large, and (124) follows due to $\text{tr}[\mathbf{X}] \leq \|\mathbf{X}\| \text{rank}(\mathbf{X})$, and we have used $\|\acute{\mathbf{C}}_{j,k}^{-1}\| \leq 1/\text{Im}(z)$, $\|\mathbf{\Upsilon}_{j,k}\| = 1$ and $\text{rank}\,(\mathbf{\Upsilon}_{j,k}) < N-K$. Additionally, from Lemma 9

$$\left| \rho_j^N - \rho_{j,k}^{N''} \right| \leq \frac{1}{\text{Im}(z)(N-K)} \quad (126)$$

So from (122), (125), (126) and

$$\left| \rho_j^N - \rho_{j,k}^N \right| \leq \left| \rho_j^N - \rho_{j,k}^{N''} \right| + \left| \rho_{j,k}^{N''} - \rho_{j,k}^{N'} \right| + \left| \rho_{j,k}^{N'} - \rho_{j,k}^N \right| \quad (127)$$

we have that $\max_{k \leq K} \left| \rho_{j,k}^N - \rho_j^N \right| \xrightarrow{a.s.} 0$ in the limit considered, as stated in (10).

We now use (10) and Lemma 6 to show (17) and (18), where $n$, $a_{N,n}$, $b_N$, and $c_{N,n}$ in the lemma correspond to $k$, $\rho_{j,k}^N$, $\rho_j^N$, and $D_{j,k}$, respectively. Checking the conditions of Lemma 6, $D_{j,k}$ is real-valued and $|\rho_j^N| \leq \text{Im}(z)^{-1}$, so it remains to show that (73) is satisfied. To do this, note that Lemma 8 may be applied to $\rho_{j,k}^N$, since $|\mathbf{v}_{j,k}| = 1$ and $\|\acute{\mathbf{C}}_{j,k}\| \leq JD_{\max} + |z| < \infty$. Therefore, $\rho_{j,k}^N \in \mathbb{C}^+$ almost surely, which establishes (73). Therefore, (74) and (75) of Lemma 6 give (17) and (18).

## APPENDIX III

### PROOFS FOR SECTION II-C.

We first show that, in the limit considered, $\max_{k \leq K} \left| \pi_{i,k}^N - \pi_i^N \right| \xrightarrow{a.s.} 0$ for $i = 1$. The proof for $i = 2$ is analogous. Define

$$\pi_{1,k}^{N'} = \frac{1}{N-K} \operatorname{tr} \left[ \mathbf{\Upsilon}_{1,k} \acute{\mathbf{D}}_2 \acute{\mathbf{V}}_2 \acute{\mathbf{B}}_{1,k}^{-1} \right] \tag{128}$$

$$\pi_{1,k}^{N''} = \frac{1}{N-K} \operatorname{tr} \left[ \mathbf{\Upsilon}_1 \mathbf{X}_2 \acute{\mathbf{B}}_{1,k}^{-1} \right] \tag{129}$$

where $\mathbf{\Upsilon}_{1,k}$ is defined in (121). Using the same steps as taken in the proof of (10) in Appendix II, it is straightforward to show that $\max_{k \leq K} \left| \pi_{1,k}^N - \pi_{1,k}^{N'} \right| \xrightarrow{a.s.} 0$, and using a realization for which this holds, that $\left| \pi_{1,k}^{N'} - \pi_{1,k}^{N''} \right| \to 0$. However, in order to show $\left| \pi_1^N - \pi_{1,k}^{N''} \right| \to 0$, we require the following lemma.

**Lemma 14** *For $N \times N$ Hermitian $\mathbf{X}_1$ and $\mathbf{X}_2$, $\mathbf{v}, \mathbf{u} \in \mathbb{C}^N$, $z \in \mathbb{C}^+$, let*

$$v_a = \mathbf{v}^\dagger \mathbf{U}_2 \mathbf{D}_2 \mathbf{A}^{-1} \mathbf{D}_2^\dagger \mathbf{U}_2^\dagger \mathbf{u} \tag{130}$$

$$v_b = \mathbf{v}^\dagger \mathbf{X}_2 \mathbf{B}^{-1} \mathbf{u} \tag{131}$$

*where $\mathbf{U}_2 \mathbf{D}_2^2 \mathbf{U}_2^\dagger$ is the s.v.d. of $\mathbf{X}_2$ such that $\mathbf{U}_2^\dagger \mathbf{U}_2 = \mathbf{I}_K$, and*

$$\mathbf{A} = \mathbf{D}_2^\dagger \mathbf{U}_2^\dagger \mathbf{X}_1 \mathbf{U}_2 \mathbf{D}_2 - z\mathbf{I}_K \tag{132}$$

$$\mathbf{B} = \mathbf{X}_1 \mathbf{X}_2 - z\mathbf{I}_N \tag{133}$$

*Then $v_a = v_b$.*

*Proof:* The proof is easily obtained using induction on the rank of $\mathbf{X}_1$, and the matrix inversion lemma. ∎

Note that we can write $\pi_{1,k}^{N''} = \frac{1}{N-K} \sum_{\ell=1}^{N-K} \acute{\mathbf{s}}_{1,\ell}^\dagger \mathbf{X}_2 \acute{\mathbf{B}}_{1,k}^{-1} \acute{\mathbf{s}}_{1,\ell}$, where $\acute{\mathbf{s}}_{1,\ell}$ is defined via $\mathbf{\Upsilon}_1 = \sum_{\ell=1}^{N-K} \acute{\mathbf{s}}_{1,\ell} \acute{\mathbf{s}}_{1,\ell}^\dagger$. Moreover, we may apply Lemma 14 to each term in this representation, and write the result as matrix trace. The same argument applies to $\pi_1^N$ (we omit the details). It can then be shown that Lemma 9 can be applied to $\left| \pi_1^N - \pi_{1,k}^{N''} \right|$ using this alternate representation. The result is

$$\left| \pi_1^N - \pi_{1,k}^{N''} \right| \leq \frac{D_{\max}}{\operatorname{Im}(z)(N-K)} \tag{134}$$

Combining the preceding results, we have that $\max_{k \leq K} \left| \pi_{1,k}^N - \pi_1^N \right| \xrightarrow{a.s.} 0$ in the limit considered, as stated in (34).

We now show (40)–(42) using Lemma 6, analogous to the proof of (17) and (18) in Appendix II. Here, $n$, $a_{N,n}$, $b_N$, and $c_{N,n}$ in the lemma correspond to $k$, $\pi_{1,k}^N$, $\pi_1^N$, and $D_{j,k}$, respectively.

In this case, we note that $\left| \pi_1^N \right| \leq D_{\max} \operatorname{Im}(z)^{-1}$, which is the required uniform upper bound on the term corresponding to $b_N$ in the lemma. In addition, in order to satisfy condition (73), note that Lemma 8 may be applied to $\pi_{1,k}^N$, after writing $\pi_{1,k}^N = \mathbf{v}_{1,k}^\dagger \acute{\mathbf{V}}_2 \acute{\mathbf{D}}_2 (\bar{\mathbf{B}}_{1,k}^\dagger \bar{\mathbf{B}}_{1,k})^{-1} \acute{\mathbf{D}}_2^\dagger \acute{\mathbf{V}}_2^\dagger \mathbf{v}_{1,k}$ using Lemma 14, where $\bar{\mathbf{B}}_{1,k} = \acute{\mathbf{D}}_2^\dagger \acute{\mathbf{V}}_2^\dagger \mathbf{X}_{1,k} \acute{\mathbf{V}}_2 \acute{\mathbf{D}}_2 - z \mathbf{I}_K$. To show that the condition (83) is satisfied in the application of Lemma 8, note that Lemma 5 implies

$$\max_{k \leq K_j} \left| \mathbf{v}_{1,k}^\dagger \mathbf{X}_2 \mathbf{v}_{1,k} - \frac{1}{N} \operatorname{tr}[\mathbf{X}_2] \right| \xrightarrow{a.s.} 0 \tag{135}$$

in the limit considered. Moreover, $\left| \frac{1}{N} \operatorname{tr}[\mathbf{X}_2] - \mathsf{E}[X_2] \right| \xrightarrow{a.s.} 0$ due to (1), and $\mathsf{E}[X_2]$ is positive due to the assumption that the distribution of $X_2$ does not have all of its mass at zero.

In summary, $\left| \pi_1^N \right|$ is uniformly bounded above, and $\pi_{1,k}^N \in \mathbb{C}^+$ almost surely due to Lemma 8, which together imply that Lemma 6 may be applied to give (40)–(42).

## APPENDIX IV

### PROOF OF THEOREM 1.

Note that since the e.d.f.'s of the eigenvalues of $\mathbf{H}_j \mathbf{H}_j^\dagger$ and $\mathbf{P}_j = \mathbf{A}_j^2$ converge in distribution almost surely to compactly supported distributions, we have

$$\lim_{N \to \infty} \frac{1}{N} \sum_{k=1}^N \mathrm{f}(D_{j,k}) = \mathsf{E}\left[\mathrm{f}(H_j)\right] \tag{136}$$

$$\lim_{K_j \to \infty} \frac{1}{K_j} \sum_{k=1}^{K_j} \mathrm{f}(P_{j,k}) = \mathsf{E}\left[\mathrm{f}(P_j)\right] \tag{137}$$

for any bounded, continuous function $\mathrm{f}$ on the support of $H_j$ and $P_j$, respectively. In order to simplify the proof which follows, we also assume that $|z| < \infty$.

We seek $\gamma^N = G_R^N(z) = \frac{1}{N} \operatorname{tr}[\mathbf{R}^{-1}]$. To this end, first consider removing column $k$ from $\mathbf{S}_j$. The matrix inversion lemma gives

$$\mathbf{R}^{-1} \mathbf{h}_{j,k} = \frac{1}{1 + P_{j,k} \rho_{j,k}^N} \mathbf{R}_{d_{j,k}} \mathbf{h}_{j,k} \tag{138}$$

where $\mathbf{R}_{d_{j,k}} = \mathbf{R} - P_{j,k} \mathbf{h}_{j,k}^\ddagger$, $\mathbf{h}_{j,k} = \mathbf{H}_j \mathbf{s}_{j,k}$, and $\rho_{j,k}^N = \mathbf{h}_{j,k}^\dagger \mathbf{R}_{d_{j,k}}^{-1} \mathbf{h}_{j,k}$.

Under the limit considered,

$$\max_{k \leq K_j} \left| \rho_{j,k}^N - \rho_j^N \right| \xrightarrow{a.s.} 0 \tag{139}$$

where

$$\rho_j^N = \begin{cases} \frac{1}{N}\mathrm{tr}[\mathbf{H}_j^\dagger \mathbf{R}^{-1}\mathbf{H}_j] & , \text{ i.i.d. } \mathbf{S}_j, \\ \frac{1}{N-K_j}\mathrm{tr}[\mathbf{\Upsilon}_j \mathbf{H}_j^\dagger \mathbf{R}^{-1}\mathbf{H}_j] & , \text{ iso. } \mathbf{S}_j, \end{cases} \quad (140)$$

and $\mathbf{\Upsilon}_j = \mathbf{I}_N - \mathbf{S}_j \mathbf{S}_j^\dagger$. This can be proven following the same steps as the proof of (10) in Appendix II, where Lemma 4 is used in place of Lemma 5 for the first step with i.i.d. $\mathbf{S}_j$, and where the bounds obtained also depend on $H_{\max}$, which is finite by assumption.

Also, note that under the limit considered,

$$\max_{k \leq K_j} \left| \mathbf{s}_{j,k}^\dagger \mathbf{H}_j^\dagger \mathbf{H}_j \mathbf{s}_{j,k} - \mathbf{E}[H_j] \right| \xrightarrow{a.s.} 0 \quad (141)$$

from the Borel-Cantelli lemma, Lemma 4, Lemma 5, and (136). Moreover, since the distribution of $H_j$ does not contain all mass at zero, $\mathbf{E}[H_j]$ is positive.

We now focus on a realization for which (136), (137), (139) and (141) holds. Applying (138) to an expansion of the identity $\mathbf{I}_N = \mathbf{R}\mathbf{R}^{-1}$, we obtain

$$1 + z\gamma^N = \frac{1}{N}\sum_{j=1}^{J}\sum_{k=1}^{K_j} P_{j,k} \mathbf{h}_{j,k}^\dagger \mathbf{R}^{-1} \mathbf{h}_{j,k} = \sum_{j=1}^{J} \frac{\alpha_j}{K_j} \sum_{k=1}^{K_j} \frac{P_{j,k}\rho_{j,k}^N}{1+P_{j,k}\rho_{j,k}^N} \quad (142)$$

Using Lemma 6, it can be shown that for any $j = 1, \ldots, J$,

$$\max_{k \leq K_j} \left| \frac{P_{j,k}\rho_{j,k}^N}{1+P_{j,k}\rho_{j,k}^N} - \frac{P_{j,k}\rho_j^N}{1+P_{j,k}\rho_j^N} \right| \to 0 \quad (143)$$

following the same steps as the proof of (17) in Appendix II. The only significant differences are that (141) is used to satisfy condition (83) of Lemma 8, and we also require the fact from [24] that $\|\mathbf{S}_j\| \xrightarrow{a.s.} 1 + \sqrt{\alpha_j}$.

Hence from (142) and (143),

$$\left| 1 + z\gamma^N - \sum_{j=1}^{J} \alpha_j \rho_j^N \mathcal{P}_j^N \right| \to 0 \quad (144)$$

$$\mathcal{P}_j^N = \frac{1}{K_j}\sum_{k=1}^{K_j} \frac{P_j}{1+P_j \rho_j^N} \quad (145)$$

Until this point, this derivation differs little from that encountered in Sections II-B and II-C.

To proceed, we need the following extension of [14, Proposition 2].

**Proposition 1** *For the model (47), the distribution of both the Stieltjés transform of the e.d.f. of the eigenvalues of $\sum_{j=1}^{J}(\mathbf{H}_j\mathbf{S}_j\mathbf{A}_j)^\ddagger$ and the MMSE SINR are invariant to the substitution of $\mathbf{V}\mathbf{D}_j$ for $\mathbf{H}_j$, where $\mathbf{V}$ is an $N \times N$ Haar-distributed random unitary matrix, and $\mathbf{D}_j$ is an $N \times N$ diagonal matrix containing the singular values of $\mathbf{H}_j$.*

*Proof:* For some $N \times N$ Haar distributed matrix $\mathbf{T}$, note that

$$\gamma^N = \frac{1}{N}\text{tr}[\mathbf{R}^{-1}] = \frac{1}{N}\text{tr}[\mathbf{T}\mathbf{T}^\dagger\mathbf{R}^{-1}] = \frac{1}{N}\text{tr}[(-z\mathbf{I}_N + \sum_{j=1}^{J}(\mathbf{T}\mathbf{H}_j\mathbf{S}_j\mathbf{A}_j)^\ddagger)^{-1}] \quad (146)$$

and

$$\rho_j^N = \frac{1}{N}\text{tr}[\mathbf{H}_j^\dagger\mathbf{R}^{-1}\mathbf{H}_j] = \frac{1}{N}\text{tr}[\mathbf{H}_j^\dagger\mathbf{T}^\dagger\mathbf{T}\mathbf{R}^{-1}\mathbf{T}^\dagger\mathbf{T}\mathbf{H}_j] \quad (147)$$

$$= \frac{1}{N}\text{tr}[(\mathbf{T}\mathbf{H}_j)^\dagger(-z\mathbf{I}_N + \sum_{j=1}^{J}(\mathbf{T}\mathbf{H}_j\mathbf{S}_j\mathbf{A}_j)^\ddagger)^{-1}(\mathbf{T}\mathbf{H}_j)] \quad (148)$$

Writing $\mathbf{T}\mathbf{H}_j\mathbf{S}_j = (\mathbf{T}\mathbf{U}_{j,1})\mathbf{D}_j(\mathbf{U}_{j,2}^\dagger\mathbf{S}_j)$, where $\mathbf{U}_{j,1}\mathbf{D}_j\mathbf{U}_{j,2}^\dagger$ is the singular value decomposition of $\mathbf{H}_j$, the unitary invariance of $\mathbf{T}$ and $\mathbf{S}_j$ gives the result. ∎

Therefore, in the remainder of this appendix, we substitute $\mathbf{H}_j$ with $\mathbf{V}\mathbf{D}_j$ everywhere. We denote the $n^{\text{th}}$ column of $\mathbf{V}$ and $\mathbf{S}_j^\dagger$ as $\mathbf{v}_n$ and $\tilde{\mathbf{s}}_{j,n}$, respectively, $1 \leq n \leq N$. Denote the diagonal elements of $\mathbf{D}_j$ as $\{d_{j,1}, \ldots, d_{j,N}\}$.

In contrast to the derivations in Section II-B, note that if $(\mathbf{H}_j\mathbf{S}_j\mathbf{A}_j)^\ddagger$, $j = 1, \ldots, J$, were free, instead of taking $\mathbf{H}_j = \mathbf{V}\mathbf{D}_j$, we would set $\mathbf{H}_j = \mathbf{V}_j\mathbf{D}_j$, where $\mathbf{V}_j$, $j = 1, \ldots, J$, are *independent* $N \times N$ Haar distributed random unitary matrices. This is the key departure point of this (non-free) derivation from the (free) derivations of Section II.

Now consider the removal of the $n^{\text{th}}$ column of $\mathbf{V}$, for some $0 < n \leq N$, i.e.,

$$\mathbf{R} = -z\mathbf{I}_N + \sum_{j=1}^{J}((\mathbf{H}_{j,t_n}\mathbf{S}_{j,t_n} + d_{j,n}\mathbf{v}_n\tilde{\mathbf{s}}_{j,n}^\dagger)\mathbf{A}_j)^\ddagger \quad (149)$$

$$= \mathbf{R}_{t_n} + \sum_{j=1}^{J} d_{j,n}\mathbf{u}_{j,n}\mathbf{v}_n^\dagger + d_{j,n}\mathbf{v}_n\mathbf{u}_{j,n}^\dagger + d_{j,n}^2 c_{j,n}\mathbf{v}_n\mathbf{v}_n^\dagger \quad (150)$$

where

$$\mathbf{R}_{t_n} = -z\mathbf{I}_N + \sum_{j=1}^{J}(\mathbf{H}_{j,t_n}\mathbf{S}_{j,t_n}\mathbf{A}_j)^\ddagger \quad (151)$$

$$\mathbf{u}_{j,n} = \mathbf{H}_{j,t_n}\mathbf{S}_{j,t_n}\mathbf{A}_j^2\tilde{\mathbf{s}}_{j,n} \quad (152)$$

$$c_{j,n} = \tilde{\mathbf{s}}_{j,n}^\dagger\mathbf{A}_j^2\tilde{\mathbf{s}}_{j,n} \quad (153)$$

and $\mathbf{H}_{j,t_n}$ and $\mathbf{S}_{j,t_n}$ denote $\mathbf{H}_j$ and $\mathbf{S}_j$ with their $n^{\text{th}}$ column and row removed, respectively.

In what follows, we shall apply Lemma 12 and Lemma 13 to (150), where $\mathbf{Y}_N$, $\mathbf{X}_N$, $\mathbf{v}_{N,j}$, $\mathbf{u}_N$, and $c_{N,j}$ in the statement of Lemma 12 correspond to $\mathbf{R}$, $\mathbf{R}_{t_n}$, $d_{j,n}\mathbf{u}_{j,n}$, $\mathbf{v}_n$, and $c_{j,n}$, respectively. We shall now verify that the conditions of the lemmas are satisfied.

Define $\tau_{j,n}^N = \mathbf{u}_{j,n}^\dagger \mathbf{R}_{t_n}^{-1} \mathbf{u}_{j,n}$. Since $\mathbf{R}_{t_n} \mathbf{v}_n = -z\mathbf{v}_n$, we have

$$\mathbf{v}_n^\dagger \mathbf{R}_{t_n}^{-1} \mathbf{v}_n = -z^{-1} \tag{154}$$

$$\mathbf{v}_n^\dagger \mathbf{R}_{t_n}^{-1} \mathbf{u}_{j,n} = 0, \tag{155}$$

and (155) implies condition (105). It can be shown that for $i \neq j$, $\mathbf{u}_{i,n}^\dagger \mathbf{R}_{t_n}^{-1} \mathbf{u}_{j,n} \xrightarrow{a.s.} 0$ in the limit, satisfying condition (106). This is clear for $\mathbf{S}_j$ and $\mathbf{S}_i$ i.i.d. from [8, Corollary 1], and for isometric $\mathbf{S}_j$ and/or $\mathbf{S}_i$ the proof requires considering $N \left| \mathbf{u}_{i,n}^\dagger \mathbf{R}_{t_n}^{-1} \mathbf{u}_{j,n} \right|^2$. Since $|\mathbf{v}_n| = 1$, condition (107) is satisfied, and finally condition (108) is satisfied by assumption.

Now, note that due to Lemma 5, Lemma 4, and the Borel-Cantelli lemma,

$$\max_{n \leq N} \left| c_{j,n} - \alpha_j \bar{p}_j^N \right| \xrightarrow{a.s.} 0 \tag{156}$$

in the limit considered, where $\bar{p}_j^N = \frac{1}{K_j} \sum_{k=1}^{K_j} P_{j,k}$.

In Appendix VI, we show that

$$\max_{n \leq N} \left| \tau_{j,n}^N - \tau_j^N \right| \xrightarrow{a.s.} 0, \tag{157}$$

$$\tau_j^N = \begin{cases} \frac{1}{N} \text{tr}[\mathbf{H}_j \mathbf{S}_j \mathbf{A}_j^4 \mathbf{S}_j^\dagger \mathbf{H}_j^\dagger \mathbf{R}^{-1}] & \text{, i.i.d. } \mathbf{S}_j, \\ \frac{1}{N} \sum_{n=1}^N \tau_{j,n}^N & \text{, iso. } \mathbf{S}_j. \end{cases} \tag{158}$$

We will now focus on a realization for which (136), (137), (156) and (157) holds.

In order to determine $\tau_j^N$, first note that

$$\frac{1}{N} \text{tr}[\mathbf{H}_j \mathbf{S}_j \mathbf{A}_j^4 \mathbf{S}_j^\dagger \mathbf{H}_j^\dagger \mathbf{R}^{-1}] = \alpha_j \frac{1}{K_j} \sum_{k=1}^{K_j} P_{j,k}^2 \mathbf{h}_{j,k}^\dagger \mathbf{R}^{-1} \mathbf{h}_{j,k} \tag{159}$$

from which we can obtain

$$\left| \frac{1}{N} \text{tr}[\mathbf{H}_j \mathbf{S}_j \mathbf{A}_j^4 \mathbf{S}_j^\dagger \mathbf{H}_j^\dagger \mathbf{R}^{-1}] - \alpha_j (\bar{p}_j^N - \mathcal{P}_j^N) \right| \to 0, \tag{160}$$

in the limit considered, in the same manner as the proof of (144).

Considering $\tau_{j,n}^N$ when $\mathbf{S}_j$ is isometric, using $\mathbf{S}_j^\dagger \mathbf{S}_j = \mathbf{I}_{K_j}$, we have

$$\frac{1}{N} \text{tr}[\mathbf{H}_j \mathbf{S}_j \mathbf{A}_j^4 \mathbf{S}_j^\dagger \mathbf{H}_j^\dagger \mathbf{R}^{-1}] = \frac{1}{N} \text{tr}[\mathbf{H}_j \mathbf{S}_j \mathbf{A}_j^2 \mathbf{S}_j^\dagger \mathbf{S}_j \mathbf{A}_j^2 \mathbf{S}_j^\dagger \mathbf{H}_j^\dagger \mathbf{R}^{-1}] \tag{161}$$

$$= \frac{1}{N} \sum_{n=1}^N \text{tr}[(\mathbf{u}_{j,n} + d_{j,n} c_{j,n} \mathbf{v}_n)(\mathbf{u}_{j,n} + d_{j,n} c_{j,n} \mathbf{v}_n)^\dagger \mathbf{R}^{-1}] \tag{162}$$

Focusing on the argument of the sum in (162), we have from Lemma 13 that

$$\left| (\mathbf{u}_{j,n} + d_{j,n} c_{j,n} \mathbf{v}_n)^\dagger \mathbf{R}^{-1} (\mathbf{u}_{j,n} + d_{j,n} c_{j,n} \mathbf{v}_n) - \left( \tau_{j,n}^N + \frac{d_{j,n}^2 (c_{j,n} - \tau_{j,n}^N)^2}{-z + \sum_i d_{i,n}^2 (c_{i,n} - \tau_{i,n}^N)} \right) \right| \to 0 \tag{163}$$

where we have used Lemma 12 in order to simplify the terms corresponding to $\acute{u}_N$, $\acute{v}_{N,j}$, $\varepsilon_{N,j}^{(1)}$, and $\varepsilon_{N,j}^{(2)}$. Using (156) and (157), it is straightforward to show that in the limit considered

$$\max_{n \leq N} \left| \frac{d_{j,n}^2 (c_{j,n} - \tau_{j,n}^N)^2}{-z + \sum_i d_{i,n}^2 (c_{i,n} - \tau_{i,n}^N)} - \frac{d_{j,n}^2 (\alpha_j \bar{p}_j^N - \tau_j^N)^2}{-z + \sum_i d_{i,n}^2 (\alpha_i \bar{p}_i^N - \tau_i^N)} \right| \to 0 \tag{164}$$

and hence combining (160), (162), (163), and (164) we obtain

$$\left| \tau_j^N - \alpha_j (\bar{p}_j^N - \mathcal{P}_j^N) + (\alpha_j \bar{p}_j^N - \tau_j^N)^2 \mathcal{H}_j^N \right| \to 0 \tag{165}$$

in the limit considered, where

$$\mathcal{H}_j^N = \frac{1}{N} \sum_{n=1}^N \frac{d_{j,n}^2}{-z + \sum_i (\alpha_i \bar{p}_i^N - \tau_i^N) d_{i,n}^2} \tag{166}$$

In order to determine $\rho_j^N$, note that

$$\frac{1}{N} \text{tr}[\mathbf{H}_j^\dagger \mathbf{R}^{-1} \mathbf{H}_j] = \frac{1}{N} \sum_{n=1}^N d_{j,n}^2 \text{tr}[\mathbf{v}_n \mathbf{v}_n^\dagger \mathbf{R}^{-1}]. \tag{167}$$

Focusing on the argument of the sum in (167), like (163), we have from Lemma 13 that

$$\max_{n \leq N} \left| \mathbf{v}_n^\dagger \mathbf{R}^{-1} \mathbf{v}_n - \frac{1}{-z + \sum_i (c_{i,n} - \tau_{i,n}^N) d_{i,n}^2} \right| \to 0 \tag{168}$$

Using (156) and (157), it is straightforward to show that in the limit considered

$$\max_{n \leq N} \left| \frac{1}{-z + \sum_i (c_{i,n} - \tau_{i,n}^N) d_{i,n}^2} - \frac{1}{-z + \sum_i (\alpha_i \bar{p}_i^N - \tau_i^N) d_{i,n}^2} \right| \to 0 \tag{169}$$

and hence combining (167), (168), and (169) we obtain

$$\left| \frac{1}{N} \text{tr}[\mathbf{H}_j^\dagger \mathbf{R}^{-1} \mathbf{H}_j] - \mathcal{H}_j^N \right| \to 0. \tag{170}$$

In addition, considering the term $\frac{1}{N} \text{tr}[\mathbf{H}_j \mathbf{S}_j \mathbf{S}_j^\dagger \mathbf{H}_j^\dagger \mathbf{R}^{-1}]$ which appears in the expansion of $\rho_j^N$ for isometric $\mathbf{S}_j$, we have

$$\left| \frac{1}{N} \text{tr}[\mathbf{H}_j \mathbf{S}_j \mathbf{S}_j^\dagger \mathbf{H}_j^\dagger \mathbf{R}^{-1}] - \alpha_j \rho_j^N (1 - \rho_j^N \mathcal{P}_j^N) \right| \to 0 \tag{171}$$

which is proven in an identical manner as (160).

Combining (140), (170) and (171), we obtain

$$\left| \rho_j^N - \mathcal{H}_j^N \right| \to 0 \quad , \text{ i.i.d. } \mathbf{S}_j, \tag{172}$$

$$\left| \rho_j^N - \frac{1}{1 - \alpha_j} \left( \mathcal{H}_j^N - \alpha_j \rho_j^N (1 - \rho_j^N \mathcal{P}_j^N) \right) \right| \to 0 \quad , \text{ iso. } \mathbf{S}_j. \tag{173}$$

It follows that for any realization for which (136), (137), (144), (160), (165), (172), and (173) hold, $\left| \gamma^N - \gamma \right| \to 0$, $\left| \rho_j^N - \rho_j \right| \to 0$, and $\left| \tau_j^N - \tau_j \right| \to 0$, where $\gamma$, $\rho_j$, and $\tau_j$, satisfy (59)–(62).

# APPENDIX V

## PROOF OF LEMMA 12.

The proof of Lemma 12 is by induction on $J$. We drop the subscript $N$ for brevity, such that $\mathbf{X}_N$, $\mathbf{Y}_N$, $\mathbf{u}_N$, $\mathbf{v}_{N,j}$, and $c_{N,j}$ are denoted $\mathbf{X}$, $\mathbf{Y}$, $\mathbf{u}$, $\mathbf{v}_j$, and $c_j$, respectively. Define $\mathbf{X}_{(0)} = \mathbf{X}$, and

$$\mathbf{X}_{(j)} = \mathbf{X}_{(j-1)} + \mathbf{v}_j \mathbf{u}^\dagger + \mathbf{u}\mathbf{v}_j^\dagger + c_j \mathbf{u}\mathbf{u}^\dagger \tag{174}$$

for $j = 1, \ldots, J$, noting that $\mathbf{Y} = \mathbf{X}_{(J)}$. The proof also depends on showing that

$$\left| \mathbf{X}_{(i)}^{-1} \mathbf{v}_j - \mathbf{X}_{(0)}^{-1} \mathbf{v}_j \right| \xrightarrow{a.s.} 0 \quad, j > i. \tag{175}$$

for $i = 1, \ldots, J$.

Clearly, for $J = 0$, (109)–(110) and (175) are true. Therefore, let us assume the hypothesis is true for all $J$ less than some fixed $I$, and consider $J = I$. Proving that (175) holds for $i = I$ under the inductive assumption can be shown as an auxiliary result in the proof of Lemma 10 in [14, Lemma 6] under the assumption (106). Now, in order to apply Lemma 10 to $\mathbf{X}_{(I)}$, we check the corresponding conditions.

- Note that $\left| \mathbf{u}^\dagger \mathbf{X}_{(I-1)}^{-1} \mathbf{v}_j \right| \leq |\mathbf{u}| \left| \mathbf{X}_{(I-1)}^{-1} \mathbf{v}_j - \mathbf{X}_{(0)}^{-1} \mathbf{v}_j \right| + |\epsilon_{N,j}|$, and hence due to (108), the induction assumption (175), and (105) we have $\left| \mathbf{u}^\dagger \mathbf{X}_{(I-1)}^{-1} \mathbf{v}_j \right| \xrightarrow{a.s.} 0$. Hence condition (90) of Lemma 10 is satisfied.
- Assumptions (107)–(108) imply conditions (91)–(92) of Lemma 10, since $\|\mathbf{X}_{(I)}\| \leq \|\mathbf{X}\| + IB^2(2+B) < \infty$ from the triangle inequality.
- From the induction assumption we have

$$\left| \mathbf{u}^\dagger \mathbf{X}_{(I-1)} \mathbf{u} - \frac{u_N}{1 + u_N \sum_{j=1}^{I-1} d_{N,j}} \right| \xrightarrow{a.s.} 0 \tag{176}$$

$$\left| \mathbf{v}_I^\dagger \mathbf{X}_{(I-1)} \mathbf{v}_I - v_{N,I} \right| \xrightarrow{a.s.} 0 \tag{177}$$

Therefore, from Lemma 10 and (176)–(177), it is straightforward to obtain

$$\left| \mathbf{Y}^{-1} \mathbf{u} - \frac{\mathbf{X}_{(I-1)}^{-1}\left(\mathbf{u} - \frac{u_N}{1+u_N \sum_{j=1}^{I-1} d_{N,j}} \mathbf{v}_I\right)}{1 + \frac{u_N}{1+u_N \sum_{j=1}^{I-1} d_{N,j}} d_{N,I}} \right| \xrightarrow{a.s.} 0 \tag{178}$$

$$\left| \mathbf{Y}^{-1} \mathbf{v}_I - \frac{\mathbf{X}_{(I-1)}^{-1}\left(-v_{N,I}\mathbf{u} + (1 + c_{N,I} \frac{u_N}{1+u_N \sum_{j=1}^{I-1} d_{N,j}})\mathbf{v}_I\right)}{1 + \frac{u_N}{1+u_N \sum_{j=1}^{I-1} d_{N,j}} d_{N,I}} \right| \xrightarrow{a.s.} 0 \tag{179}$$

The result is obtained from (178)–(179) after substituting the expressions for $\mathbf{X}_{(I-1)}^{-1}\mathbf{u}$ and $\mathbf{X}_{(I-1)}^{-1}\mathbf{v}_I$ obtained via induction, and simplifying.

## APPENDIX VI

### PROOF OF $\max_{n \leq N} \left| \tau_{j,n}^N - \tau_j^N \right| \xrightarrow{a.s.} 0$

### A. i.i.d. $\mathbf{S}_j$

Define

$$\tau_{j,n}^{N'} = \frac{1}{N} \mathrm{tr}[\mathbf{A}_j^2 \mathbf{S}_{j,t_n}^\dagger \mathbf{H}_{j,t_n}^\dagger \mathbf{R}_{t_n}^{-1} \mathbf{H}_{j,t_n} \mathbf{S}_{j,t_n} \mathbf{A}_j^2] \tag{180}$$

$$\tau_{j,n}^{N''} = \frac{1}{N} \mathrm{tr}[\mathbf{A}_j^2 \mathbf{S}_{j,t_n}^\dagger \mathbf{H}_{j,t_n}^\dagger \mathbf{R}^{-1} \mathbf{H}_{j,t_n} \mathbf{S}_{j,t_n} \mathbf{A}_j^2] \tag{181}$$

As with (122), we have $\max_{n \leq N} |\tau_{j,n}^N - \tau_{j,n}^{N'}| \to 0$ almost surely from Lemma 4 and the Borel-Cantelli lemma. Now, $\mathbf{u}_{j,n}^\dagger \mathbf{R}_{t_n}^{-1} \mathbf{H}_{j,t_n} \mathbf{S}_{j,t_n} \mathbf{A}_j^4 \mathbf{S}_{j,t_n}^\dagger \mathbf{H}_{j,t_n}^\dagger \mathbf{R}_{t_n}^{-1} \mathbf{v}_n = 0$ in analogy with (155), $\sup_N \|\mathbf{H}_{t_n} \mathbf{S}_{t_n} \mathbf{A}^4 \mathbf{S}_{t_n}^\dagger \mathbf{H}_{t_n}^\dagger\| < \infty$, and

$$\left| \tau_{j,n}^{N'} - \tau_{j,n}^{N''} \right| = \frac{1}{N} \left| \mathrm{tr}[\mathbf{H}_{j,t_n} \mathbf{S}_{j,t_n} \mathbf{A}_j^4 \mathbf{S}_{j,t_n}^\dagger \mathbf{H}_{j,t_n}^\dagger (\mathbf{R}^{-1} - \mathbf{R}_{t_n}^{-1})] \right|. \tag{182}$$

So, Lemma 13 applies to (182), from which we obtain $\max_{n \leq N} \left| \tau_{j,n}^{N'} - \tau_{j,n}^{N''} \right| \to 0$.

Also,

$$\left| \tau_{j,n}^{N''} - \tau_j^N \right| = \left| \frac{1}{N} \mathrm{tr}[\mathbf{R}^{-1}(\mathbf{H}_{j,t_n} \mathbf{S}_{j,t_n} \mathbf{A}_j^4 \mathbf{S}_{j,t_n}^\dagger \mathbf{H}_{j,t_n}^\dagger - \mathbf{H}_j \mathbf{S}_j \mathbf{A}_j^4 \mathbf{S}_j^\dagger \mathbf{H}_j^\dagger)] \right|$$

$$= \left| \frac{1}{N} \mathrm{tr}[\mathbf{R}^{-1}(d_{j,n} \mathbf{u}_{j,n} \mathbf{v}_n^\dagger + d_{j,n} \mathbf{v}_n \mathbf{u}_{j,n}^\dagger + d_n^2 c_{j,n} \mathbf{v}_n \mathbf{v}_n^\dagger)] \right|$$

$$\leq \frac{1}{N} \left( 2 |d_{j,n}| \left| \mathbf{R}^{-1} \mathbf{u}_{j,n} \right| + \left| d_{j,n}^2 \right| |c_{j,n}| \left| \mathbf{R}^{-1} \mathbf{v}_n \right| \right) \tag{183}$$

From our assumptions and the application of Lemma 12, it is clear that the terms inside the bracket of (183) are uniformly bounded (i.e. independent of $n$ and $N$), so $\max_{n \leq N} \left| \tau_{j,n}^{N''} - \tau_j^N \right| \to 0$ as well.

Combining the preceding results with the triangle inequality, we have $\max_{n \leq N} \left| \tau_{j,n}^N - \tau_j^N \right| \xrightarrow{a.s.} 0$.

### B. Isometric $\mathbf{S}_j$

Let $m, n \in \{1, \ldots, N\}$ with $m \neq n$, and consider some $j \in \{1, \ldots, J\}$. Define

$$\tau_{j,n} = \mathbf{u}_{j,n}^\dagger \mathbf{R}_{t_n}^{-1} \mathbf{u}_{j,n} \tag{184}$$

$$\tau_{j,n,m} = \mathbf{u}_{j,n,m}^\dagger \mathbf{R}_{t_n}^{-1} \mathbf{u}_{j,n,m} \tag{185}$$

$$\tau'_{j,n,m} = \mathbf{u}_{j,n,m}^\dagger \mathbf{R}_{t_{n,m}}^{-1} \mathbf{u}_{j,n,m} \tag{186}$$

$$\tau''_{j,n,m} = \acute{\mathbf{u}}_{j,n,m}^\dagger \mathbf{R}_{t_{n,m}}^{-1} \acute{\mathbf{u}}_{j,n,m} \tag{187}$$

where

$$\mathbf{u}_{j,n,m} = \mathbf{H}_{j,t_{n,m}} \mathbf{S}_{j,t_{n,m}} \mathbf{A}_j^2 \tilde{\mathbf{s}}_{j,n} = \mathbf{u}_{j,n} - d_{j,m} \tilde{\mathbf{s}}_{j,m}^\dagger \mathbf{A}_j^2 \tilde{\mathbf{s}}_{j,n} \mathbf{v}_m \tag{188}$$

$$\acute{\mathbf{u}}_{j,n,m} = \mathbf{H}_{j,t_{n,m}} \mathbf{S}_{j,t_{n,m}} \mathbf{A}_j^2 \mathbf{E}_{K_j} \boldsymbol{\Theta}_j^\dagger \boldsymbol{\Psi}_{n,m} \mathbf{e}_n \tag{189}$$

$$\boldsymbol{\Psi}_{n,m} = \mathbf{e}_n \mathbf{e}_m^\dagger + \mathbf{e}_m \mathbf{e}_n^\dagger + \sum_{\ell \neq m,n}^N \mathbf{e}_\ell \mathbf{e}_\ell^\dagger \tag{190}$$

and $\mathbf{e}_n$ is an $N \times 1$ vector which contains zeros except for a 1 in the $n^{\text{th}}$ row. Note that $\tilde{\mathbf{s}}_{j,n}$ may be written as $\mathbf{E}_{K_j} \boldsymbol{\omega}_{j,n}$, where $\mathbf{E}_{K_j} = [\mathbf{I}_{K_j}, \mathbf{0}_{K_j, N-K_j}]$, and $\boldsymbol{\omega}_{j,n}^\dagger$ is the $n^{\text{th}}$ row of the $N \times N$ Haar matrix $\boldsymbol{\Theta}_j$ from which $\mathbf{S}_j$ is taken, i.e., $\mathbf{S}_j = \boldsymbol{\Theta}_j \mathbf{E}_{K_j}^\dagger$ and $\boldsymbol{\omega}_{j,n} = \boldsymbol{\Theta}_j^\dagger \mathbf{e}_n$.

Note that

$$\max_{j,m,n,(m \neq n)} \left| \tilde{\mathbf{s}}_{j,m}^\dagger \mathbf{A}_j^2 \tilde{\mathbf{s}}_{j,n} \right| \xrightarrow{a.s.} 0 \tag{191}$$

$$\max_{j,m,n,(m \neq n)} \left| \mathbf{u}_{j,m,n}^\dagger \mathbf{R}_{t_{n,m}}^{-1} \mathbf{u}_{j,n,m} \right| \xrightarrow{a.s.} 0 \tag{192}$$

$$\mathbf{u}_{j,n,m}^\dagger \mathbf{R}_{t_{n,m}}^{-1} \mathbf{v}_m^\dagger = 0, \quad \forall\, m,n \in \{1, \ldots, N\},\, m \neq n \tag{193}$$

where (191) and (192) can be shown using standard arguments after writing $\tilde{\mathbf{s}}_{j,m}$ and $\tilde{\mathbf{s}}_{j,n}$ as just described, and (193) is shown in the same way as (155). We now focus on a realization for which (191) and (192) hold.

Now, $\max_{m,n,(m \neq n)} |\tau_{j,n} - \tau_{j,n,m}| \to 0$ follows from $|\tau_{j,n} - \tau_{j,n,m}| \leq 2 \left| \tilde{\mathbf{s}}_{j,m}^\dagger \mathbf{A}_j^2 \tilde{\mathbf{s}}_{j,n} \right| |d_{j,m}| \left| \mathbf{R}_{t_n}^{-1} \mathbf{u}_{j,n} \right|$, (191), and the fact that the latter two terms are uniformly bounded.

Writing $\tau_{j,n,m} = \operatorname{tr}[\mathbf{u}_{j,n,m} \mathbf{u}_{j,n,m}^\dagger \mathbf{R}_{t_n}^{-1}]$ and similarly for $\tau'_{j,n,m}$, we have from Lemma 13 that $\max_{m,n,(m \neq n)} |\tau_{j,n,m} - \tau'_{j,n,m}| \to 0$, since the terms corresponding to $\acute{u}_N$, $\acute{v}_{N,i}$, $\varepsilon_{N,j}^{(1)}$, and $\varepsilon_{N,j}^{(2)}$ in the statement of the lemma converge to zero (independently of $m$ and $n$) due to (192) and (193).

Finally, since $\boldsymbol{\Theta}_j$ is unitarily invariant, and $\boldsymbol{\Psi}_{n,m}$ is unitary ($\boldsymbol{\Psi}_{n,m}$ is simply the permutation matrix which swaps the $n^{\text{th}}$ and $m^{\text{th}}$ entries), $\max_{m,n,(m \neq n)} |\tau'_{j,n,m} - \tau''_{j,n,m}| \to 0$.

Combining the above results gives $\max_{m,n,(m \neq n)} |\tau_{j,n} - \tau_{j,m}| \to 0$ and moreover

$$\left| \tau_j^N - \tau_{j,n}^N \right| \leq \frac{1}{N} \sum_{m=1}^N \left| \tau_{j,m}^N - \tau_{j,n}^N \right| \leq \max_{m,n,(m \neq n)} |\tau_{j,n} - \tau_{j,m}| \to 0 \tag{194}$$